\documentclass[journal=jacsat,manuscript=article]{achemso}
\usepackage[T1]{fontenc} 
\usepackage{amsmath,amssymb}
\usepackage{xcolor,soul}

\providecommand{\red}{\color{red}}

\usepackage{lipsum}
\usepackage{atbegshi}
\AtBeginDocument{\AtBeginShipoutNext{\AtBeginShipoutDiscard}}

\providecommand{\kwords}[1]{\textbf{\textit{Key words:}} #1}
\usepackage[font=small,labelfont=bf]{caption}

\title[]{Deterministic fabrication of graphene hexagonal boron nitride moir\'e superlattices}

\author{Rupini V. Kamat}
\affiliation[Cofirst]
{These authors contributed equally to this work}
\alsoaffiliation[PHYS]
{Department of Physics, Stanford University, 382 Via Pueblo Mall, Stanford, CA 94305, USA}
\alsoaffiliation[SIMES]
{Stanford Institute for Materials and Energy Science, SLAC National Accelerator Laboratory, 2575 Sand Hill Road, Menlo Park, California 94025, USA}
\author{Aaron L. Sharpe}
\affiliation[Cofirst]
{These authors contributed equally to this work}
\alsoaffiliation[PHYS]
{Department of Physics, Stanford University, 382 Via Pueblo Mall, Stanford, CA 94305, USA}
\alsoaffiliation[SIMES]
{Stanford Institute for Materials and Energy Science, SLAC National Accelerator Laboratory, 2575 Sand Hill Road, Menlo Park, California 94025, USA}
\email{aaron.sharpe@stanford.edu}
\author{Mihir Pendharkar}
\affiliation[matsci]
{Department of Materials Science and Engineering, Stanford University, 496 Lomita Mall, Stanford, CA 94305, USA}
\alsoaffiliation[SIMES]
{Stanford Institute for Materials and Energy Science, SLAC National Accelerator Laboratory, 2575 Sand Hill Road, Menlo Park, California 94025, USA}
\author{Jenny Hu}
\affiliation[PHYS]
{Department of Applied Physics, Stanford University, 348 Via Pueblo Mall, Stanford, CA 94305, USA}
\author{Steven J. Tran}
\affiliation[PHYS]
{Department of Physics, Stanford University, 382 Via Pueblo Mall, Stanford, CA 94305, USA}
\alsoaffiliation[SIMES]
{Stanford Institute for Materials and Energy Science, SLAC National Accelerator Laboratory, 2575 Sand Hill Road, Menlo Park, California 94025, USA}
\author{Gregory Zaborski Jr.}
\affiliation[matsci]
{Department of Materials Science and Engineering, Stanford University, 496 Lomita Mall, Stanford, CA 94305, USA}
\author{Marisa Hocking}
\affiliation[matsci]{Department of Materials Science and Engineering, Stanford University, 496 Lomita Mall, Stanford, CA 94305, USA}
\alsoaffiliation[SIMES]
{Stanford Institute for Materials and Energy Science, SLAC National Accelerator Laboratory, 2575 Sand Hill Road, Menlo Park, California 94025, USA}
\author{Joe Finney}
\affiliation[PHYS]
{Department of Physics, Stanford University, 382 Via Pueblo Mall, Stanford, CA 94305, USA}
\alsoaffiliation[SIMES]
{Stanford Institute for Materials and Energy Science, SLAC National Accelerator Laboratory, 2575 Sand Hill Road, Menlo Park, California 94025, USA}
\alsoaffiliation[SIMES]
{Stanford Institute for Materials and Energy Science, SLAC National Accelerator Laboratory, 2575 Sand Hill Road, Menlo Park, California 94025, USA}
\author{Kenji Watanabe}
\affiliation[NIMS0]{Research Center for Functional Materials,
National Institute for Materials Science, 1-1 Namiki, Tsukuba 305-0044, Japan}
\author {Takashi Taniguchi}
\affiliation[NIMS1]
{International Center for Materials Nanoarchitectonics,
National Institute for Materials Science,  1-1 Namiki, Tsukuba 305-0044, Japan}
\author{Marc A. Kastner}
\affiliation[PHYS]
{Department of Physics, Stanford University, 382 Via Pueblo Mall, Stanford, CA 94305, USA}
\alsoaffiliation[SIMES]
{Stanford Institute for Materials and Energy Science, SLAC National Accelerator Laboratory, 2575 Sand Hill Road, Menlo Park, California 94025, USA}
\alsoaffiliation[MIT]
{Department of Physics, Massachusetts Institute of Technology, 77 Massachusetts Avenue, Cambridge, MA 02139, USA}
\author{Andrew J. Mannix}
\affiliation[matsci]
{Department of Materials Science and Engineering, Stanford University, 496 Lomita Mall, Stanford, CA 94305, USA}
\alsoaffiliation[SIMES]
{Stanford Institute for Materials and Energy Science, SLAC National Accelerator Laboratory, 2575 Sand Hill Road, Menlo Park, California 94025, USA}
\author{Tony Heinz}
\affiliation[PHYS]
{Department of Physics, Stanford University, 382 Via Pueblo Mall, Stanford, CA 94305, USA}
\alsoaffiliation[SIMES]
{Stanford Institute for Materials and Energy Science, SLAC National Accelerator Laboratory, 2575 Sand Hill Road, Menlo Park, California 94025, USA}
\author{David Goldhaber-Gordon}
\affiliation[PHYS]
{Department of Physics, Stanford University, 382 Via Pueblo Mall, Stanford, CA 94305, USA}
\alsoaffiliation[SIMES]
{Stanford Institute for Materials and Energy Science, SLAC National Accelerator Laboratory, 2575 Sand Hill Road, Menlo Park, California 94025, USA}
\email{goldhaber-gordon@stanford.edu}

\begin{document}
\pagenumbering{arabic}

\begin{abstract}
The electronic properties of moir\'e heterostructures depend sensitively on the relative orientation between layers of the stack. 
For example, near-magic-angle twisted bilayer graphene (TBG) commonly shows superconductivity, yet a TBG sample with one of the graphene layers rotationally aligned to a hexagonal Boron Nitride (hBN) cladding layer provided the first experimental observation of orbital ferromagnetism. 
To create samples with aligned graphene/hBN, researchers often align edges of exfoliated flakes that appear straight in optical micrographs. 
However, graphene or hBN can cleave along either zig-zag or armchair lattice directions, introducing a $30^\circ$ ambiguity in the relative orientation of two flakes. 
By characterizing the crystal lattice orientation of exfoliated flakes prior to stacking using Raman and second-harmonic generation for graphene and hBN, respectively, we unambiguously align monolayer graphene to hBN at a near-$0^\circ$, not $30^\circ$, relative twist angle. 
We confirm this alignment by torsional force microscopy (TFM) of the graphene/hBN moir\'e on an open-face stack, and then by cryogenic transport measurements, after full encapsulation with a second, non-aligned hBN layer. 
This work demonstrates a key step toward systematically exploring the effects of the relative twist angle between dissimilar materials within moir\'e heterostructures.

\end{abstract}
 
{\red \kwords{moir\'e superlattice, graphene, boron nitride}}


With their reliably-calculable band structure and high degree of tunability, moir\'e heterostructures have demonstrated a great variety of strongly correlated and topological phases. 
Experimentally realizing these exotic phases requires precisely controlling the relative orientation of constituent layers of the heterostructure. 
In moir\'es where each layer is the same material, such as the canonical twisted bilayer graphene (TBG), angular control to within $\sim0.1^{\circ}$ of a targeted angle is enabled by `tear and stack' or `cut and stack' techniques, in which one crystalline exfoliated flake is bisected and the two sections are then stacked with the desired interlayer twist angle~\cite{kim_van_2016, cao_superlattice-induced_2016}. This technique is important since TBG with interlayer twist within $0.1^\circ$ of the $1.1^{\circ}$ ``magic angle'' typically features correlated states at integer fillings of the moir\'e flat bands~\cite{cao_correlated_2018}, and superconductivity at nearby fillings.~\cite{cao_unconventional_2018, yankowitz_tuning_2019, lu_superconductors_2019, stepanov_untying_2020}.

Moir\'e patterns can also form between dissimilar but isostructural materials. For such heterobilayers, alignment can have a strong effect on the structure's electronic properties, yet precisely setting the interlayer twist angle is challenging. In cut-and-stacked homobilayers, the two layers are known to begin with the same orientation before an intentional twist is introduced, but in heterobilayers such a starting zero-twist reference is not available without additional characterization.

Consider the materials pair of graphene and hexagonal boron nitride (hBN), whose lattice spacings differ by only $\sim1.8\%$. Alignment to hBN can dramatically change the electronic properties of graphene-based stacks, whether monolayer graphene\cite{moon_electronic_2014, yankowitz_emergence_2012, lee_ballistic_2016, ponomarenko_cloning_2013, dean_hofstadters_2013, hunt_massive_2013, yu_hierarchy_2014}, Bernal bilayer graphene\cite{yao_enhanced_2021, jat_higher_2024}, twisted bilayer graphene~\cite{sharpe_emergent_2019,  serlin_intrinsic_2020}, or multilayer rhombohedral graphene~\cite{chen_evidence_2019, chen_signatures_2019, chen_tunable_2020, lu_fractional_2024},.  
In this work, we are particularly motivated by the impact of hBN alignment on ground states of TBG near the magic angle. Of the many near-magic-angle TBG devices that have been fabricated and thoroughly characterized worldwide over the past five years, only two, to our knowledge, have instead demonstrated a robust orbital ferromagnetic state at 3 electrons per moir\'e unit cell at zero magnetic field~\cite{sharpe_emergent_2019, serlin_intrinsic_2020}. In each of these devices one of the hBN cladding layers was aligned to within 1$^\circ$ with the proximate graphene layer.
Of the two TBG devices, one exhibited quantized Hall resistance at zero magnetic field and $n/n_s = 3$~\cite{serlin_intrinsic_2020}, indicating that this ferromagnetic state is a quantum anomalous Hall (QAH) state.

What causes QAH in TBG, and why has it been so rarely observed? The nearly-flat conduction and valence minibands of TBG are connected by Dirac points at the corners of the mini Brillouin zone, which are protected by $C_{2z}T$, where $C_{2z}$ is $xy$-inversion symmetry and $T$ is time reversal symmetry. 
Achieving QAH requires breaking both.
Breaking $C_{2z}$ gaps the Dirac points, and breaking $T$ then allows uneven filling of the fourfold spin- and valley-degenerate copies of the moir\'e flat band. 
Each copy has non-zero Chern number $\pm 1$, so filling an odd number of copies gives a net Chern number and thus a quantum-Hall-like state. 

Initial theories for QAH in near-magic-angle TBG implied that the hBN-graphene alignment observed in the two ferromagnetic TBG devices was necessary and sufficient to break $C_{2z}T$~\cite{zhang_twisted_2019, bultinck_mechanism_2020}: hBN lacks $C_{2z}$ symmetry so aligning monolayer graphene to hBN breaks $C_{2z}$\cite{kindermann_zero-energy_2012, yankowitz_emergence_2012} in the graphene, while flat band electron-electron interactions spontaneously break $T$. However these theories did not account for the graphene-hBN lattice mismatch, which causes the inversion symmetry in graphene to be broken only locally at periodically-spaced AA stacking sites in the graphene-hBN moir\'e. Subsequent theory of QAH in near-magic angle TBG suggests a more stringent condition on the structure: commensurability between the co-existing TBG and graphene-hBN moir\'es~\cite{shi_moire_2021, cea_band_2020, mao_quasiperiodicity_2021}. 
The simplest commensurability occurs when the two moir\'es share the same period and orientation, so that the entire three-layer system forms a single moir\'e pattern. 
This criterion is satisfied only at specific pairs of twist angles, notably a graphene-hBN twist angle of $\pm0.6^\circ$ and a graphene-graphene twist angle of $\pm1.2^\circ$\cite{shi_moire_2021, lai_imaging_2023}.

Recent STM studies suggest that local graphene-graphene-hBN commensurability occurs over a broader range of twist angles than predicted for rigid lattices, indicating that it may be energetically favorable~\cite{lai_imaging_2023}. Still, in practice we are unlikely to realize perfectly commensurate moir\'e structures globally in such a 3-layer system. Fortunately, theory suggests that QAH should be observable by transport even in devices with one or both twist angles up to $\sim 0.1^\circ$ off from the ideal (in which case a ``supermoir\'e'' or ``moir\'e of moir\'es'' is formed)~\cite{shi_moire_2021}.

In short, precise, reliable, and verifiable control over the relative angle between graphene and hBN is necessary (and likely sufficient) for reproducing and further investigating QAH in TBG. It should also allow the exploration of novel states in other twist-controlled moir\'e heterostructures built from graphene and hBN, or from a wider range of layered materials including transition metal dichalcogenides. To date, rotationally aligning graphene with other isostructural van der Waals materials has relied on visually aligning long ($\mathcal{O}(10\, \mathrm{\mu m})$ or longer) `straight' edges of exfoliated materials. 
These edges often (but not always) result from cleavage along high-symmetry planes\cite{you_edge_2008, neubeck_direct_2010} of the  crystal lattice, and thus can serve as a proxy for the crystallographic orientation of the flakes\cite{hunt_massive_2013}.
The crystallinity of an edge is often corroborated by the presence of multiple straight edges differing in orientation by integer multiples of $30^\circ$. 
However, such cleavage can occur along two distinct crystallographic directions, yielding straight edges with zigzag or armchair termination.
Though simulations and experiments on suspended graphene membranes have suggested that graphene shows a preference for ripping along armchair edges~\cite{kawai_self-redirection_2009, kim_ripping_2012}, in practice both types of edges are frequently observed in exfoliated flakes of graphene and hBN.  
Thus, visually aligning straight edges in hBN and graphene can accidentally match a zigzag graphene edge with an armchair hBN edge or vice versa, resulting in 30$^\circ$ misalignment. Furthermore, even seemingly straight edges can be non-crystallographic and contain some mixture of zigzag and armchair structure~\cite{canifmmode_mboxcelse_cfiado_influence_2004, casiraghi_raman_2009}.

Here, we outline a general approach for fabrication and rapid verification of Van der Waals heterostructures with defined relative twist angle. We demonstrate this method on graphene/hBN structures, but it should work quite broadly for other pairs of exfoliatable materials.
Our process flow is a generalization of what is already used for heterobilayers made from two different transition metal dichalcogenides (TMDs), where second-harmonic generation optical spectroscopy is already well-established as a guide for setting the relative orientation of the two layers~\cite{rivera_observation_2015, rivera_valley-polarized_2016, seyler_signatures_2019, shabani_deep_2021}. 
We use second harmonic generation spectroscopy on hBN flakes and Raman spectroscopy on graphene flakes to
(1) confirm whether straight edges of graphene and hBN flakes are crystallographic, and 
(2) identify crystallographic edges as either zig-zag or armchair, eliminating ambiguity in how exfoliated graphene-hBN flakes should be oriented during stacking to form a moir\'e superlattice.  
We then use torsional force microscopy (TFM)~\cite{pendharkar_torsional_2024} not only to verify near-alignment of graphene and hBN, but also to serve as a transport-independent real-space probe of moir\'e structural parameters.
Finally, we use low-temperature transport measurements to verify the existence of the graphene-hBN superlattice and extract a moir\'e unit cell area consistent with the TFM measurements performed prior to encapsulation of the graphene.
In sum, these techniques enable stacking dissimilar materials at near-aligned twist angle and rapidly confirming the existence of an associated moir\'e at intermediate fabrication steps

\section{Results}

We first use second harmonic generation spectroscopy to characterize the orientation of hBN exfoliated flakes. In any non-centrosymmetric crystal, illumination with a laser at frequency $f$ causes emission at 2$f$, a phenomenon known as second-harmonic generation (SHG). The strength of the second harmonic signal depends on the laser polarization relative to the lattice directions of the crystal, so measuring the polarization dependence allows determining the crystal orientation. Samples that are non-centrosymmetric and thus susceptible to this method for determining orientation include atomic monolayers or flakes of odd number of layers of TMDs or hBN~\cite{li_probing_2013, kim_second-harmonic_2019}, as well as Bernal-stacked trilayer graphene~\cite{shan_stacking_2018}. This method is commonly used when preparing to stack two different TMD monolayers to form a heterobilayer with a specific twist angle ~\cite{rivera_observation_2015, rivera_valley-polarized_2016, seyler_signatures_2019, shabani_deep_2021}.
More salient for our purposes, a small but measurable SHG signal has also been reported in {\em many-layer} hBN flakes, regardless of layer number parity: non-negligible thickness of the hBN flakes compared to the laser wavelength (1060 nm in our case) causes a gradient in the electric field strength, breaking c-axis inversion symmetry in the light-sample interaction and thus allowing quadrupole contributions to the SHG signal~\cite{ponath_chapter_1991, yao_enhanced_2021}.

For hBN, the orientation dependence is described by a 6-fold symmetric pattern $I_{||} = I_0 \cos^2(3\theta)$, where $\theta$ is the angle between the laser polarization $\vec{P}$ and a mirror plane of the crystal, and $I_{||}$ is the component of the second harmonic signal polarized parallel to the pump laser polarization~\cite{li_probing_2013}. 

\begin{figure}
    \centering
    \includegraphics[width = 0.8\linewidth]{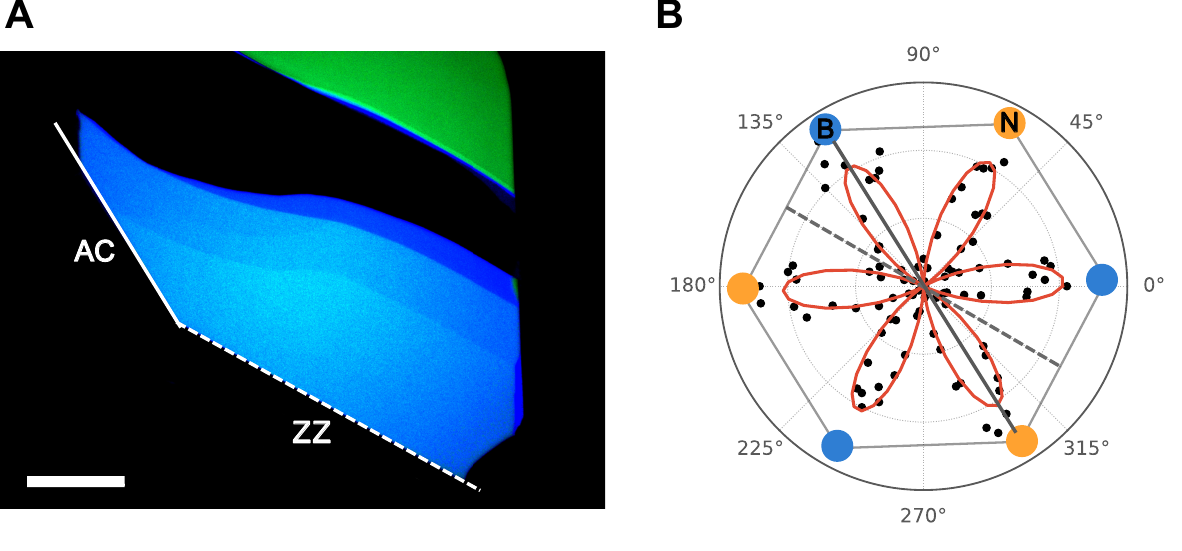}
    \caption{\textbf{Polarization-resolved SHG of many-layer hBN} 
    \textbf{A)} Optical image of an hBN flake. 
    Solid and dashed white lines indicate straight edges of interest. Scale bar is 20 $\mu$m.  
    \textbf{B)} SHG signal intensity as a function of laser polarization. 
    The straight edge indicated by the dashed (solid) white line in the optical image lies along a node (maximum) in the polarization-resolved SHG data marked by a dashed (solid) grey line, indicating a zigzag (armchair) edge. 
    }
    \label{fig:SHG}
\end{figure}

Consistent with prior reports~\cite{yao_enhanced_2021}, we observed SHG and the expected 6-fold dependence on polarization in all 15+ many-layer exfoliated hBN flakes of 30 to 60 nm thickness we studied — see Figure~\ref{fig:SHG} for one example — though roughly half were presumably even-layered and thus centrosymmetric; see Figure~\ref{fig:SHG} for one such example. 
The orientation of minima and maxima in the SHG intensity as a function of polarization are used to identify straight edges of the flake as either armchair or zigzag. 
Armchair edges in hBN are parallel to a mirror plane of the lattice, whereas zigzag edges are oriented between mirror planes of the lattice.
Thus, a crystalline edge for which $\vec{P}$ parallel to the edge yields a node (maximum) in the SHG signal is a zigzag (armchair) edge. Edges which do not line up with either a node or a maximum in the SHG signal are non-crystallographic, and hence not useful for establishing orientation.

Additional SHG measurements are shown in the Supplement (Figure \ref{fig:SuppSHG}). As noted above, SHG signals have also been observed in Bernal ABA-stacked trilayer graphene, which is non-centrosymmetric\cite{shan_stacking_2018}. 
Thus, SHG could be used to determine crystal orientation for some graphene-based structures, in preparation for stacking with defined twist angle. 
However, {\em monolayer} graphene, which we intend to align with hBN in a stack, is centrosymmetric and so does not produce a SHG signal. Its orientation must be characterized another way.

\begin{figure}
    \centering
    \includegraphics[width = 0.8\linewidth]{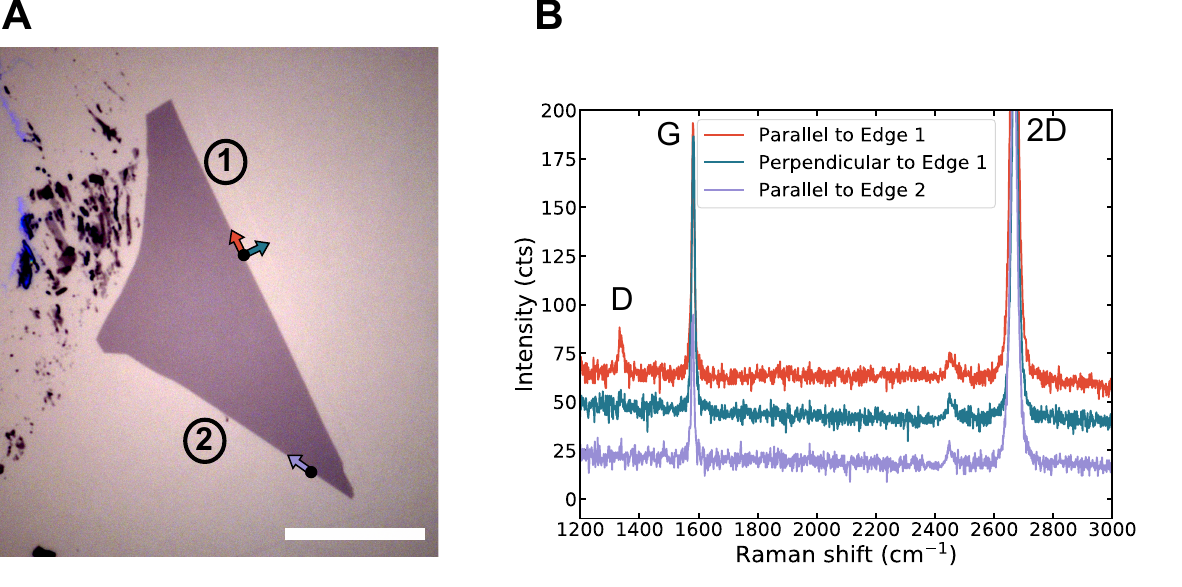}
    \caption{\textbf{Polarized Raman on monolayer graphene} 
    \textbf{A)} Optical image of an exfoliated graphene flake with straight edges, separated by a 30$^\circ$ corner. 
    Scale bar is 20 $\mu$m. 
    \textbf{B)} Raman spectra taken at the  edges of the flake. 
    Location of laser spot and direction of laser polarization for each spectrum is indicated by the solid black points and colored arrows in A. The D peak on Edge 1 that vanishes when the laser is polarized perpendicular to the edge indicates that edge 1 of A is likely armchair. Edge 2 is offset from edge 1 by 30$^\circ$, and shows no D peak signal, so is likely zigzag.
    }
    \label{fig:PRaman}
\end{figure}

To determine the crystallographic nature of straight edges in exfoliated graphene flakes, we use polarized Raman spectroscopy.
The D peak at $\sim$1350 cm$^{-1}$ in Raman spectroscopy of graphene (Figure \ref{fig:PRaman}B) is absent in the interior of a pristine graphene flake like those produced by exfoliation from a high-quality graphite crystal. This peak originates from a two-step intervalley scattering process~\cite{canifmmode_mboxcelse_cfiado_influence_2004}: inelastic scattering of the excited electron (or hole) with a phonon, then elastic scattering off of a feature that breaks translation symmetry, such as a point defect or sample edge (Supplement Figure \ref{fig:ZZAC}A). 
Because a uniform edge only transfers momentum along its normal vector, a perfect zigzag edge cannot scatter electrons between different valleys, whereas an armchair edge efficiently induces intervalley scattering (Supplement Figure \ref{fig:ZZAC}B). 
Thus, in an exfoliated monolayer graphene flake we expect to observe a D peak only near an armchair edge (or an edge containing armchair segments). 
Even for an armchair edge, the intensity of the D peak strongly depends on the excitation laser polarization~\cite{canifmmode_mboxcelse_cfiado_influence_2004, you_edge_2008, casiraghi_raman_2009, gruneis_inhomogeneous_2003}:maximal for laser polarization parallel to the edge, minimal — zero for an ideal edge — for polarization perpendicular to the edge. The ratio of D peak intensities for the two orientations can provide a measure of edge disorder, such as microscopic segments of differing termination.~\cite{casiraghi_raman_2009}. 

On a single crystal graphene flake, if two edges differ in orientation by an odd integer multiple of 30$^\circ$ (Figure \ref{fig:PRaman}A), and each edge has well-defined termination, then one edge must be zigzag and the other armchair.
Indeed, in Raman spectra taken at each edge with the laser polarization oriented parallel to the given edge, we observe a clear D peak at the right edge and no D peak on the bottom left edge (Figure \ref{fig:PRaman}B). 
Rotating the laser polarization to be perpendicular to the right edge, we see that the D peak disappears. 
This indicates that the right edge (Edge 1) has armchair edge termination, and the left edge (Edge 2) has zigzag edge termination.

Similar characterization performed on other graphene flakes can be seen in the Supplement (Figures \ref{fig:SuppRamanSuccess} and \ref{fig:SuppRamanUnsuccess}). In ideal cases, as exemplified by flakes in Figures \ref{fig:PRaman}, \ref{fig:Stack2Praman}, and \ref{fig:SuppRamanSuccess}, particular edges can be unambiguously assigned as zigzag or armchair. However in some flakes (three out of the eight flakes measured for this work), edges separated by an odd integer multiple of 30$^\circ$ both exhibit a D peak (Figure \ref{fig:SuppRamanUnsuccess}). This can occur when an edge either is not crystallographic or is oriented macroscopically along a zigzag axis but contains microscopic armchair segments, preventing unambiguous assignment of edge termination. To target specific relative alignment of graphene and hBN in a stack, we use only graphene flakes whose edge terminations we can definitively assign.

Having directly identified the crystallographic termination of straight edges on a graphene flake and an hBN flake, we can fabricate aligned graphene-hBN heterostructures by visually aligning straight edges of now-known termination during stacking, offsetting by 30$^\circ$ if we are pairing a zigzag edge with an armchair one.
Using standard dry transfer techniques (see Methods), we first pick up the hBN flake shown in Figure~\ref{fig:SHG}A. 
We then rotationally align the zigzag edge of the hBN flake (now on the stamp) with the zigzag edge of the graphene flake in Figure~\ref{fig:PRaman}A, and pick up the graphene flake. With the aligned graphene-hBN heterostructure now on the polymer stamp, we use torsional force microscopy (TFM) to characterize the moir\'e.

TFM is a local probe technique which uses a torsional resonance mode of an AFM cantilever to probe changes in local dynamic friction, reliably imaging open-face graphene-graphene and graphene-hBN moir\'es~\cite{pendharkar_torsional_2024}. 
The technique is non-destructive and can be applied to a stack still mounted on a polymer stamp.
After TFM imaging, encapsulation can be completed, the stack dropped onto a substrate, and standard lithography used to pattern a device for transport measurements.

\begin{figure}
    \centering
    \includegraphics[width = 0.8\linewidth]{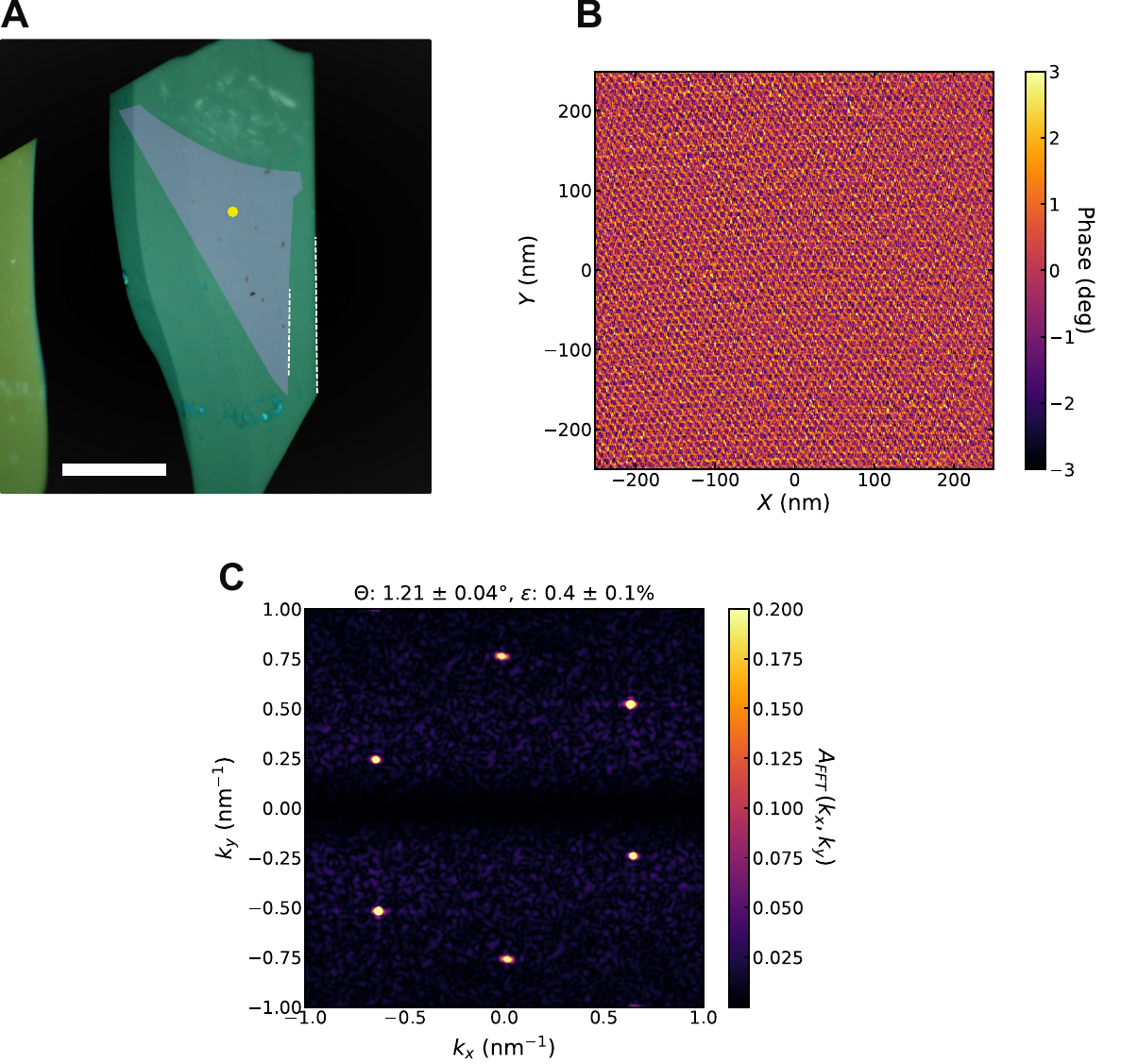}
    \caption{\textbf{Torsional Force Microscopy of graphene-hBN moir\'e}
    \textbf{A)} Optical image of the aligned graphene-hBN stack on a polymer stamp. In preparation for stacking, the zigzag edge of the graphene flake from Figure \ref{fig:PRaman}A (false colored in purple) was visually aligned with the zigzag edge of the hBN flake from Figure \ref{fig:SHG}.
    Dashed white lines indicate the zigzag edges of each flake. Yellow dot indicates rough location of TFM scan.
    Scale bar is 20 $\mu$m. 
    \textbf{B)} Phase component of the TFM signal over a 500 nm x 500 nm scan. 
    \textbf{C)} FFT of TFM scan in B. Moir\'e parameters are extracted from FFT peak positions by fitting to a simple model with twist angle ($\theta$) and uniaxial heterostrain ($\epsilon$) as fit parameters. 
    }
    \label{fig:TFM}
\end{figure}

A 500 nm x 500 nm TFM scan on our aligned graphene-hBN stack shows a clear moir\'e superlattice in the phase component of the TFM signal (Figure \ref{fig:TFM}). 
A Fourier transform of these data, yields a set of sharp peaks associated with the periodicities of the moir\'e superlattice. 
Using a simple model with twist angle and uniaxial heterostrain as fit parameters, we extract from the peak positions an estimated graphene-hBN misalignment of $1.21 \pm 0.04 ^\circ$ and uniaxial heterostrain of $0.4 \pm 0.1 \%$ (Figure~\ref{fig:TFM}B). 
We do not correct for thermal drift and piezo actuator creep in the TFM measurement, which can distort the image of the moir\'e superlattice and lead to an overestimation of strain but should not as strongly influence the extracted twist angle. 
These effects will be examined more closely in upcoming work.

Using the same process flow described here, we fabricated a second graphene-hBN heterostructure, shown in the Supplement. 
For that stack, TFM images yield a graphene-hBN twist angle of $\sim1.9^\circ$, again (as intended) close to $0^\circ$ rather than $30^\circ$, demonstrating that the spectroscopic characterization correctly identifies the crystallographic axes of both graphene and hBN.

Following TFM characterization, the graphene-hBN stack is deposited onto an annealed hBN-graphite gate heterostructure (previously assembled and deposited onto a SiO$_2$/Si substrate), and patterned into two Hall bars (see Methods). 
The bottom hBN flake is deliberately misaligned with respect to the aligned graphene-hBN, to ensure that no moir\'e is formed with the bottom hBN.

\begin{figure*}
    \centering
    \includegraphics[width = 0.8\linewidth]{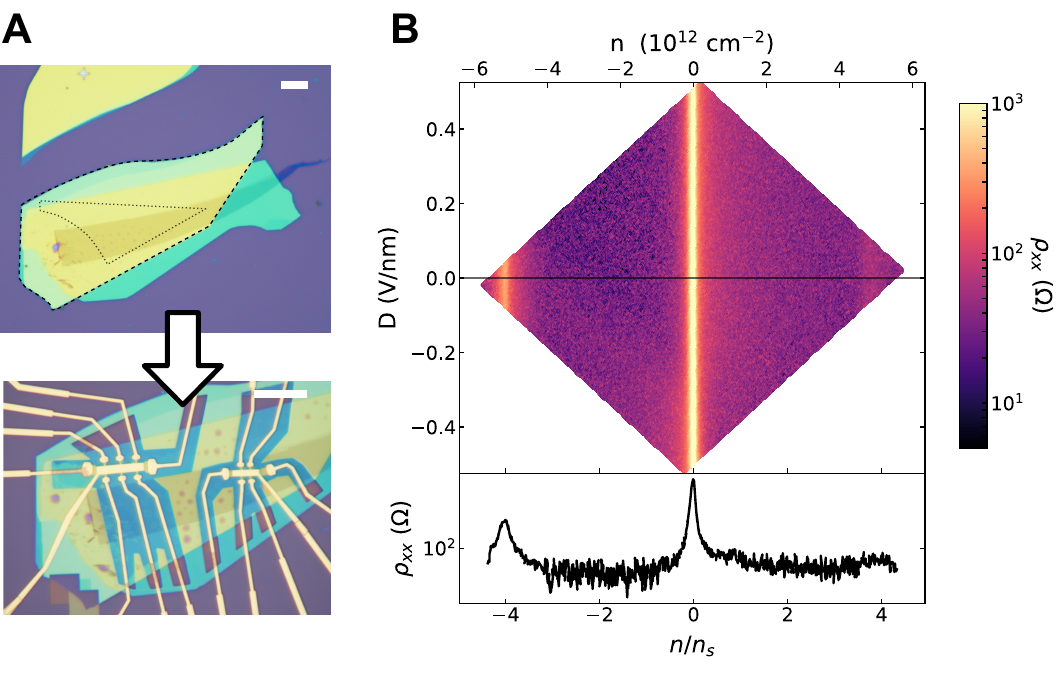}
    \caption{\textbf{Transport characterization of graphene-hBN moir\'e at 1.5K} 
    \textbf{A)} Top optical image shows the hBN-encapsulated monolayer graphene heterostructure prior to patterning. 
    Dashed black outlines indicate the borders of the pre-characterized graphene and hBN flakes; the hBN flake being the larger of the two.
    The stack is patterned into two Hall bars (bottom optical image), one with a doped Si back gate (left) and the other with a graphite back gate (right). 
    Each Hall bar has a metal top gate. Scale bars are 20 $\mu$m. 
    \textbf{B)} Top - Color map of longitudinal resistivity in the lefthand Hall bar of A, as a function of carrier density $n$ and displacement field $D$. 
    Bottom -  Line cut of resistance vs carrier density along $D = 0$. The resistance peak at $n = -5.15e12$ cm$^{-2}$ indicates full emptying of the graphene-hBN moir\'e superlattice ($n/n_s = -4$). 
    }
    \label{fig:Transport}
\end{figure*}

Graphene aligned to hBN exhibits peaks in resistivity not only at charge neutrality but also at fillings of 4 holes and/or 4 electrons per moir\'e unit cell. 
Here we observe such a peak at density $n = -5.15E12 \pm 0.05E12$ cm$^{-2}$. 
Associating this with 4 holes per moir\'e unit cell yields a unit cell area 77.7 $\pm$ 0.8 nm$^2$, corresponding to a graphene-hBN twist angle of $1.10^\circ \pm 0.01^\circ$. 
The error bars here are dominated by uncertainty in gate capacitance, which is calibrated by fitting the slopes of features in the Landau fan diagram (Figure \ref{fig:Transport}C). 
The locations of gaps in density and magnetic field follow the Diophantine relation: for integer $s$ and $t$, $n/n_s = t (\phi/\phi_0) + s$, where $n_s$ is the carrier density corresponding to 4 electrons per moir\'e unit cell, $\phi$ is the magnetic flux per moir\'e unit cell, and $\phi_0=h/e$ is the magnetic flux quantum.
In this sample, gaps we observe correspond to ($s = 0, -4$) and ($t = \pm 2, 6, 10, 14,...$).

\begin{figure}
    \centering
    \includegraphics[width = 0.8\linewidth]{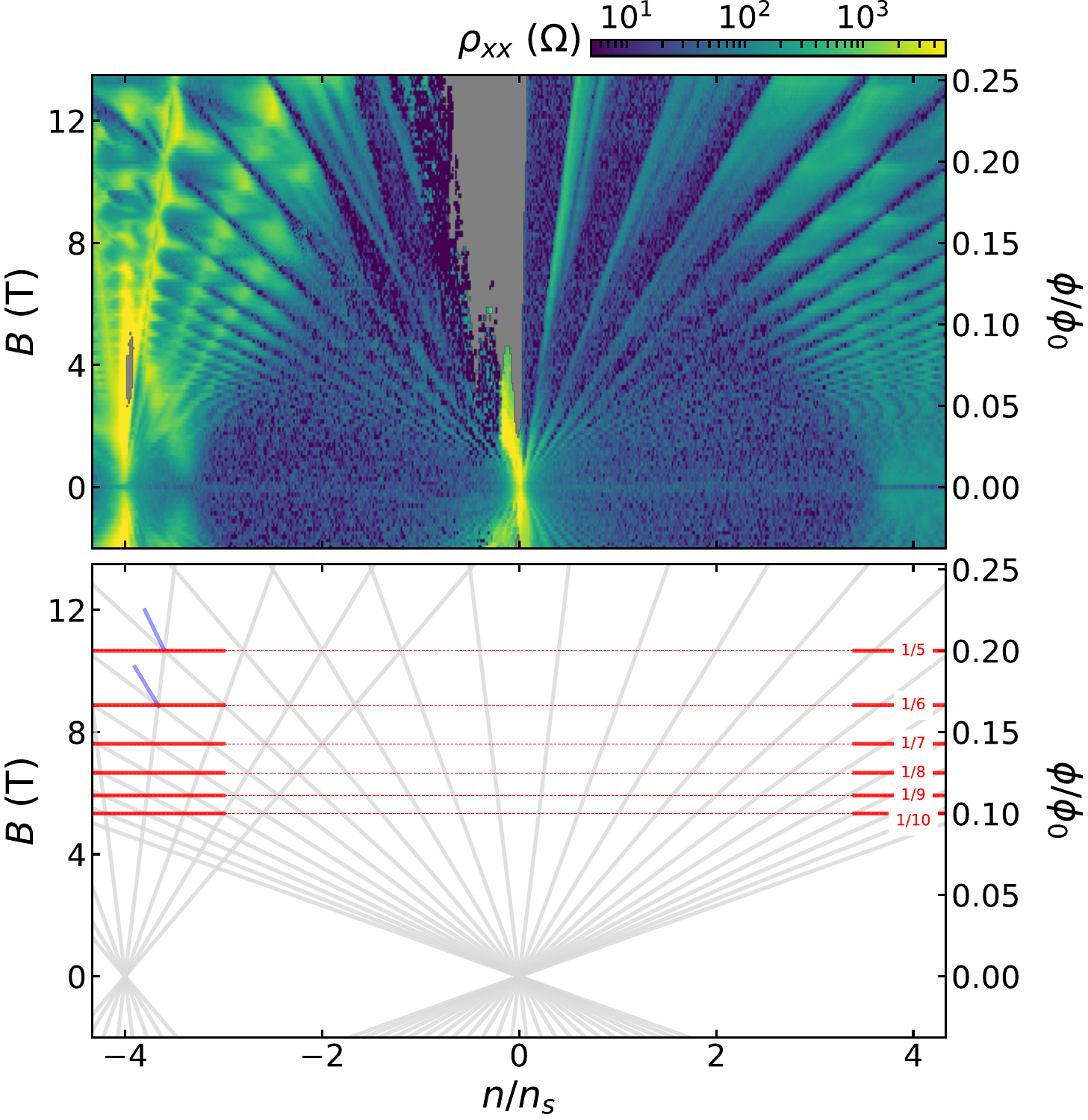}
    \caption{\textbf{Wannier diagram of graphene-hBN moir\'e at 1.5K} 
    Top - Landau fan of longitudinal resistivity at $D=0$, for the same Hall bar as in Fig. 4B.
    Bottom - Hofstadter energy spectrum features observed in the Landau fan measurement. Gaps in the fractal Hofstadter energy spectrum emanating from charge neutrality ($s = 0$, $t =  \pm 2, 6, 10, 14, ...$) and from the graphene-hBN superlattice miniband edge ($s = -4$, $t =  \pm 2, 6, 10$) are indicated by grey lines. 
    At magnetic fields where these gaps intersect ($\phi/\phi_0 = 1/m$ for $m = 5, 6, 7, 8, 9$) are local minima in $\rho_{xx}$, marked by red lines (clearest at higher $n/n_s$). 
    Indications of broken-symmetry states with $(s, t) = (-2, -8)$ and $(-2, -10)$ are marked by blue lines. 
    }
    \label{fig:LandauFan}
\end{figure}

We also measure clear Brown-Zak oscillations~\cite{krishna_kumar_high-temperature_2017} (Figure \ref{fig:LandauFan}). 
Though these carrier-density-independent features can in principle occur at any simple fraction, we measure them at $\phi / \phi_0 = 1/m$ for integer $m$, where gaps from different $s$ intersect.
The magnetic field values at which these features occur are a direct measure of moir\'e unit cell area, independent of gate capacitance. 
In this sample, we thereby extract a unit cell area of 77.7 $\pm$ 0.4 nm$^2$, corresponding to a graphene-hBN twist angle of $1.10^\circ \pm 0.06$. 
Here, error bars are set by the width of the oscillations in magnetic field.

Measurements of the second Hall bar produced from this same heterostructure are shown in the Supplement. 
The average moir\'e cell area in the second Hall bar is 58.0 nm$^2$, corresponding to a graphene-hBN twist angle of 1.4$^\circ$. 
This indicates a substantial variation (0.3$^\circ$) in the graphene-hBN twist between different locations in the heterostructure.
Spatial variation in twist angle is generically observed in moir\'e heterostructures. Variation this large may appear somewhat surprising in an annealed graphene/hBN moir\'e, a type of sample which is typically assumed to be particularly robust against twist angle disorder. However, this stack contained many bubbles (see Supplement Figure \ref{fig:Stack1AFM}). 
Though each Hall bar was defined in a relatively bubble-free region, the two Hall bars were separated by bubbles, giving an opportunity for twist angle to vary between the Hall bars. 
The TFM measurements only covered a 500 nm x 500nm area, and so did not capture the full range of moir\'e structure present across an entire Hall bar, let alone both Hall bars. Still, the TFM-extracted twist angle ($\sim$1.2$^\circ$) is consistent with the transport-extracted twist angle range of 1.1$^\circ$ to 1.4$^\circ$.
This indicates that the moir\'e superlattice likely did not change dramatically during encapsulation, device fabrication, or cryogenic cooldown. This validates in-process TFM as a tool to determine superlattice period, select regions of relatively uniform period for device fabrication, and inform analysis of transport measurements on completed encapsulated devices. 

We have combined multiple techniques to enable reliable formation and validation of low-twist-angle graphene-hBN moir\'es. First, we used optical spectroscopy techniques to identify the crystallographic orientation of both graphene and hBN, guiding the rotation angle chosen to match the orientation of the two flakes as they are stacked. After stacking these two flakes but before encapsulation with a second hBN layer, we used TFM to verify the existence of a moir\'e and characterize its period. Edge assignment determined through SHG on hBN was also independently confirmed through atomic-resolution TFM (see Supplement Figure \ref{fig:AR-TFM}). After encapsulation and nanofabrication of Hall bars, we performed cryogenic transport measurements and found that the moir\'e unit cell area extracted from transport measurements agrees reasonably well with that measured by TFM prior to encapsulation and nanofabrication.

The same edge pre-characterization and stacking process was followed for a second open-faced graphene-hBN aligned stack described in detail in the Supplement. 
TFM measurements on this stack showed a graphene-hBN twist angle of 1.9$^\circ$. 
Transport measurements were not performed on this stack.
The fact that both heterostructures fabricated showed graphene-hBN alignment close to 0$^\circ$, not 30$^\circ$, indicates that the likelihood of accidental alignment is relatively low. 

There are some limitations to this process. 
First, the techniques we use are effective for binary assignment of straight edges as either zigzag or armchair, but are not usable on flakes lacking straight edges or flakes whose apparently straight edges are disordered/non-crystallographic. 
The latter seems common in graphene, where we see a number of flakes with apparently straight edges that produce an inconclusive D-peak signal (Supplement Figure \ref{fig:SuppRamanUnsuccess}). 
This does not limit our reliability in correctly aligning graphene to hBN when we stack the two together -- we simply do not proceed with stacking of graphene flakes featuring such apparently non-crystallographic edges.
Still, in our experience with exfoliation, straight edges are less common on monolayer graphene flakes than on tens-of-nm-thick hBN flakes, and the prevalence of non-crystallographic/disordered straight edges on graphene further lowers the proportion of graphene flakes which are suitable for alignment with hBN. In the future, ambiguity in graphene edge termination might be addressed with atomic-lattice-resolution AFM-based measurements such as TFM~\cite{pendharkar_torsional_2024} or conductive AFM on an appropriate substrate~\cite{sumaiya_true_2022}.
Graphene flakes of different thicknesses are often found attached to each other or closely spaced following exfoliation, in which case the orientation has been found to be preserved (or nearly so) between flakes~\cite{hu_controlled_2023}. Should monolayer graphene happen to be found attached to or near a thicker flake, optical spectroscopy (e.g. SHG on Bernal trilayer graphene\cite{shan_stacking_2018}) or atomic-resolution TFM of the thicker region could thus be used to infer orientation of the monolayer. 

Another limitation is a lack of precise control over the final graphene-hBN twist angle. 
In the device presented above, we targeted a 0$^\circ$ twist angle during stacking, but TFM and transport measurements indicate a twist angle ranging from $1.1^\circ - 1.4^\circ$ in different regions of the heterostructure. 
These measured twist angles agree with the angle between zigzag edges in AFM images of the stack (Supplement Figure \ref{fig:Stack1AFM}), to the accuracy with which we can extract the angle from those AFM images.
This indicates that the observed misalignment is not from inaccurate characterization of lattice orientation, but instead from imprecise setting of initial alignment and/or shifting of flakes during dry transfer pickup. This is reminiscent of our experience fabricating TBG using tear- or cut-and-stack methods, where, despite guaranteed initial alignment between the two layers, the final observed twist angle in transport often differs from the intended angle by tenths of a degrees or even more.

The degree of alignment we achieve so far is not sufficient for systematically exploring electronic phases of hBN-aligned TBG. A proposed criterion for observing a QAH state in such a system is close proximity to a specific pair of commensurate angles\cite{shi_moire_2021, cea_band_2020, mao_quasiperiodicity_2021}), demanding setting both the graphene-graphene angle and the graphene-hBN angle to within 0.1$^\circ$. Reaching this benchmark will require significant improvement in our initial alignment and/or stacking techniques, which will be the subject of further work. 

\section{Discussion}
Though work remains to improve the accuracy of the target angle, as explained above, we have taken major steps toward more reliable and repeatable fabrication of graphene-hBN moir\`e superlattices.
We have also integrated TFM measurements as an accurate and non-invasive technique for characterizing the graphene-hBN twist angle mid-device fabrication. 
This in-process characterization provides two important advantages: (1) Filtering out stacks which do not have the intended twist angle before time is invested in lithography and transport measurements, and (2) Serving as a transport-independent probe of device structure, including local information on not just twist angle but also heterostrain, enabling more systematic study of the connection between structural parameters and low-energy electronic phases. 

\section{Methods}
Graphene and hBN flakes are prepared by standard mechanical exfoliation with Scotch magic tape onto a 300nm SiO$_2$/Si substrate, annealed at 90-110 C for several minutes.

Polarized Raman measurements are taken on a Horiba XploRA+ Confocal Raman system using a 532nm excitation laser and a 2400 gr/mm grating. The laser is focused through a 100x objective lens (0.9 numerical aperture) for a nominal laser spot size of 376nm. Raman maps shown in the Supplement are performed with a step size of 0.4$\mu$m. Default laser polarization of the tool is oriented in the sample plane, along the vertical axis. Alignment of graphene flake edges with laser polarization is achieved by rotating the sample with the use of a precise, 360$^\circ$ manual rotation stage. The laser polarization can be rotated 90$^\circ$ by a half-wave plate, enabling efficient acquisition of polarized Raman spectra both parallel and perpendicular to a given graphene edge.

SHG measurements are performed on a home-built SHG setup. The excitation laser is a  femtosecond pulsed laser (NKT Origami Onefive 10) with wavelength 1030 nm and pulse duration <200 fs.
Collection is done with an Andor iXon Ultra EMCCD, which measures the component of the material's SHG response polarized parallel to the excitation laser.The excitation laser polarization is rotated from 0$^\circ$ to 180$^\circ$ by a Union Optics Super Achromatic half wave plate. The half wave plate has its own non-negligible SHG signal, visible in raw data shown in Supplement, which is removed via background subtraction when fitting the SHG response of hBN flakes. 

The aligned graphene-on-hBN stack in the main text is prepared using the standard dry transfer technique for assembling vdW structures. A glass slide with a thin Poly(Bisphenol A carbonate) film over a gel (Gel-Pak DGL-17-X8) stamp is brought into contact with the previously exfoliated and SHG-characterized hBN flake, which is heated to $\sim$ 80$^\circ$C. After successfully picking up of the hBN flake, the hBN flake is lowered into contact with a pre-characterized exfoliated graphene flake, with careful alignment of the zigzag edges of both flakes as determined by SHG and polarized Raman measurements. 

Moir\'e morphology characterization was performed on the open-faced graphene-hBN-stamp assembly in a Bruker Dimension Icon with a Nanoscope 5 controller. Measurements are performed with an Adama Innovations AD-2.8-SS conductive diamond tip on a PF-TUNA cantilever holder in torsional excitation mode. Procedures are described in extensive detail in~\cite{pendharkar_torsional_2024}. 

hBN above a graphite backgate is separately assembled with dry-transfer techniques, deposited onto a 300nm SiO$_2$/doped Si substrate, and annealed for 3 hrs at 500$^\circ$C in an Ar/O$_2$ atmosphere to remove polymer residue. The aligned graphene-hBN stack is then deposited on this graphite backgate stack, encapsulating the graphene. In this step, the graphene is deliberately misaligned with the bottom hBN. The final heterostructure is then washed in solvent and again annealed for 500$^\circ$C in an Ar/O$_2$ atmosphere before being patterned into two Hall bars, one using the graphite backgate and one using the doped Si substrate as a backgate, for transport measurements. A lithographically-defined Ti/Au topgate layer is deposited for both Hall bars, and the mesa is etched with a CHF$_3$/O$_2$ etch (50sccm/5sccm) before depositing Cr/Au one-dimensional edge contacts.

Transport measurements are performed at a base temperature of 1.5K in a CRYO Industries Variable Temperature Insert (VTI) with a 14T Oxford Instruments magnet. 

\begin{acknowledgement}
    We thank Zuocheng Zhang and Feng Wang for their aid in fabricating the devices and fruitful discussions. 
    We further acknowledge fruitful discussions with Xueqi Chen and Matthew Yankowitz.
 
    {\bf Funding:} Sample preparation, measurements, and analysis were supported by the US Department of Energy, Office of Science, Basic Energy Sciences, Materials Sciences and Engineering Division, under Contract DE-AC02-76SF00515. Development of tools for robotic stacking of 2D materials was supported by SLAC National Accelerator Laboratory under the Q-BALMS Laboratory Directed Research and Development funds. Device fabrication was performed at the Stanford Nano Shared Facilities which is supported by the NSF under award ECCS-2026822. D.G.-G. acknowledges support for supplies from the Ross M. Brown Family Foundation and from the Gordon and Betty Moore Foundation’s EPiQS Initiative through grant GBMF9460. K.W. and T.T. acknowledge support from the JSPS KAKENHI (Grant Numbers 21H05233 and 23H02052) and World Premier International Research Center Initiative, MEXT, Japan. J.H. acknowledges supports from NTT Research Fellowship. G.Z.Jr. acknowledges support from National Science Foundation-Graduate Research Fellowship Program, Pat Tillman Foundation, and Diversifying Academia, Recruiting Excellence doctoral fellowship program. 
    
    {\bf Author contributions:} A.S., R.K., and D.G.-G. conceived the project. A.S. and R.K. fabricated devices. K.W. and T.T. provided the hBN crystals used for fabrication. G.Z., J.F., M.P. and S.T. aided in performing and analyzing TFM measurements. J.H. and M.H. aided in performing SHG measurements. A.S. and R.K. performed Raman measurements. A.S. performed transport measurements. A.S. and R.K. analyzed the data and discussed the interpretation. M.K., A.M., T.H., and D.G.-G. supervised the experiments and analysis. The manuscript was prepared by A.S. and R.K. with input from all authors.
    
    {\bf Competing interests:} The authors declare no competing financial interests.
    
    {\bf Data and materials availability:} The data from this study are available at the Stanford Digital Repository \cite{data_repo}.
\end{acknowledgement}

\begin{suppinfo}
\setcounter{figure}{0}
\renewcommand{\theequation}{S\arabic{equation}}
\renewcommand{\thefigure}{S\arabic{figure}}

\section{Secondary Hall Bar Transport Measurements}

The second of the two Hall bars shown in Figure \ref{fig:Transport}A was also measured in the VTI at 1.5K. The transport measurements of this second device also show signatures of a graphene-hBN moir\'e superlattice (Figure \ref{fig:SuppTransport}).

\begin{figure}
        \centering
    \includegraphics[width = \linewidth]{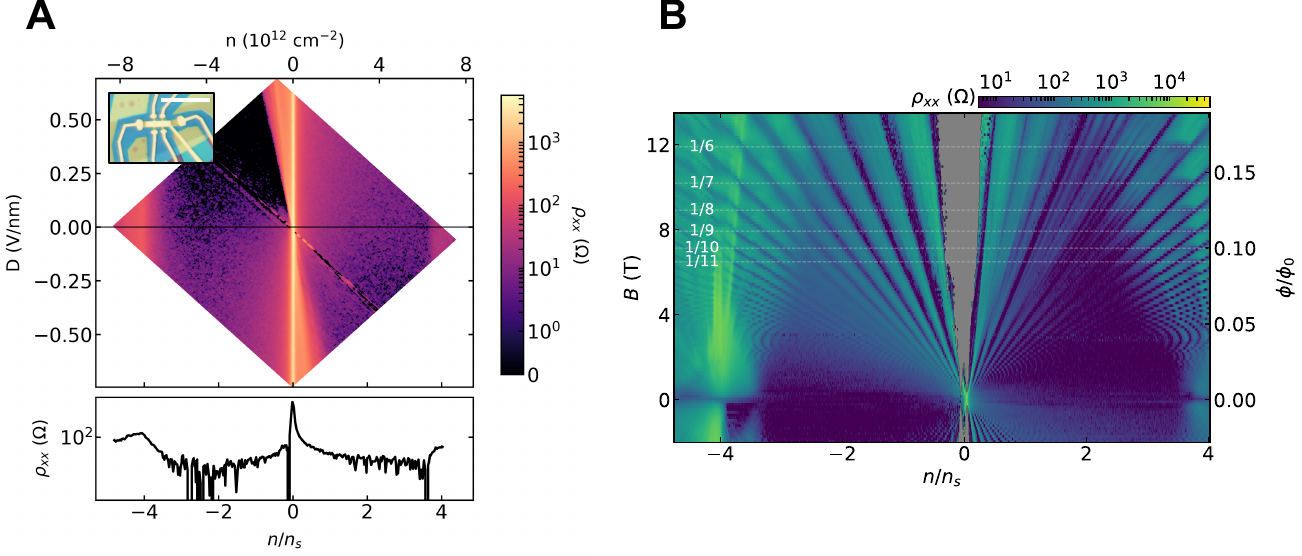}
    \caption{\textbf{Device 2 Transport Results} 
    \textbf{A)} Longitudinal resistivity ($\rho_{xx}$) as a function of carrier density $n$ and displacement field $D$. A broad peak at $n\sim6.9e12$ cm$^2$ corresponds to four holes per moir\'e unit cell ($n/n_s = -4)$.
    \textbf{B)} Longitudinal resistivity as a function of normalized carrier density $n/n_s$ and magnetic field $B$. In addition to Hofstadter energy spectrum gaps corresponding to $(s = 0$ , $t = \pm 2, 6, 10, 14...)$  and (faintly) $(s = -4$, $t = \pm 2)$, there are horizontal features corresponding to $\phi/\phi_0 = 1/m$ for $m = 6, 7, 8, 9, 10, 11$, which provide a measure of the moir\'e unit cell area. 
    } 
    \label{fig:SuppTransport}
\end{figure}

These signatures include a weak but present resistance feature at finite hole density, corresponding to $n/n_s = -4$, diagonal Landau fan features emanating from $n/n_s = -4$, and Brown-Zak oscillations with a $\propto 1/B$ periodicity. 

However, the transport features of this second device differ in detail from those shown in Figure 5. First, the resistance peak corresponding to four holes per graphene-hBN moire unit cell is weaker and broader and occurs at a higher density than the one observed in the main text (roughly 6.9E12 cm$^{-2}$ carrier density versus 5.15e12 cm$^{-2}$ carrier density). Second, the Brown-Zak oscillations corresponding to $\phi/\phi_0 = 1/m$ ($m$ being some integer) also occur at magnetic field values different from those in the main text. Both of these indicate a difference in average moire unit cell area between the two devices, with an estimated twist angle difference of roughly 0.3$^\circ$ (1.1$^\circ$ average twist angle in device 1 and 1.4$^\circ$ average twist angle in device 2). 

This level of twist variation over a length scale of tens of microns is not surprising, but this difference in transport behavior from two devices fabricated on a shared graphene-hBN heterostructure underlies ongoing challenges in using Van der Waal materials for systematic and reproducible study of novel electron transport. The long-distance variation of the graphene-hBN superlattice across the heterostructure was not explored; TFM measurements covered only a 500 nm x 500 nm area (Figure \ref{fig:TFM}). Though more time consuming, large area TFM scans can be used in the future to both inform device placement and provide more detailed structural information to clarify transport behavior as a function of position in micron scale devices.

\section{Atomic-Resolution Torsional Force Microscopy of hBN}
Atomic-resolution TFM was also performed on the hBN flake displayed in main text Figure \ref{fig:SHG}, after assembly of the fully encapsulated heterostructure as well as fabrication of the two Hall bar devices. The atomic scale TFM reveals a triangular lattice of "bright spots" which we interpret as the B-N atomic bonds. 
The FFT of the TFM image thus reveals the orientation of the atomic lattice. 
The orientation of the atomic lattice in the TFM scan independently confirms the edge assignment determined through polarization-resolved SHG measurements on the same hBN flake. 

\begin{figure}
    \centering
    \includegraphics[width = 0.9\linewidth]{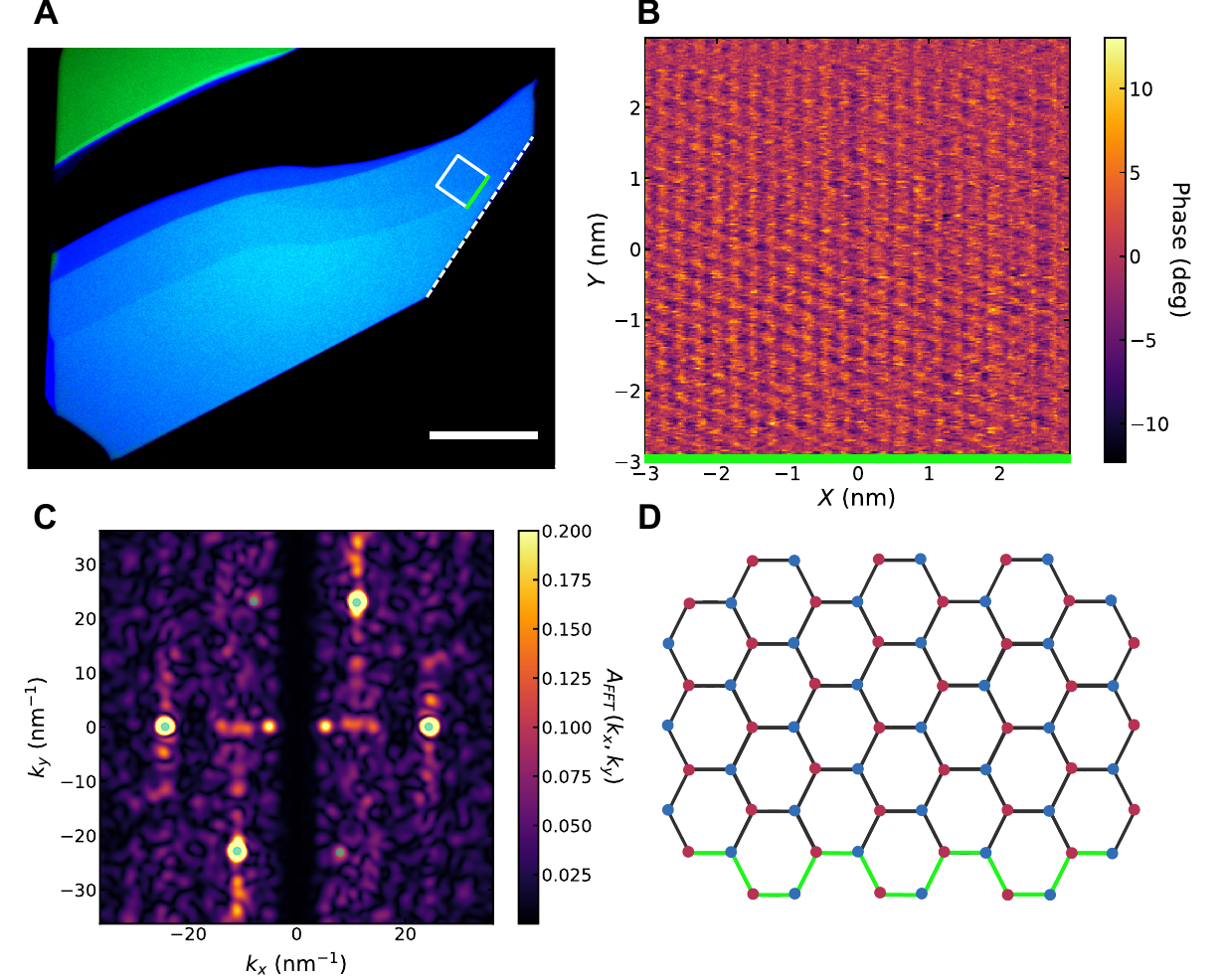}
    \caption{\textbf{Atomic-resolution TFM of hBN flake} 
    \textbf{A)} Optical image of hBN flake prior to stacking and device fabrication. White square shows the rough location and orientation of the TFM scan, with the size being greatly exaggerated and the green line indicating the bottom of the TFM scan. Scale bar is 20 $\mu$m.
    \textbf{B)} 6 nm $\times$ 6 nm TFM scan shows the hBN atomic lattice.
    \textbf{C)} FFT of the TFM scan in \textbf{B)}. Main peaks are identified with light blue spots.
    \textbf{D)} Expected atomic lattice orientation using FFT peaks identified in \textbf{C}, which confirms that the edge indicated by the solid white line in \textbf{A} has an armchair termination.
    }
    \label{fig:AR-TFM}
\end{figure}

\section{Atomic Force Microscopy of Stack 1}

After the full heterostructure in the main text was assembled, AFM measurements were taken in a Park XE 70 instrument to determine hBN thickness and to visually inspect post-stacking alignment of the pre-characterized graphene and hBN flakes (Figure \ref{fig:Stack1AFM}). The final stack had several bubbles, but the zigzag edges of the pre-characterized graphene and hBN flakes could still be identified and their relative orientation roughly measured as $1.2^\circ \pm 0.1^\circ$.

\begin{figure}
    \centering
    \includegraphics[width=450pt]{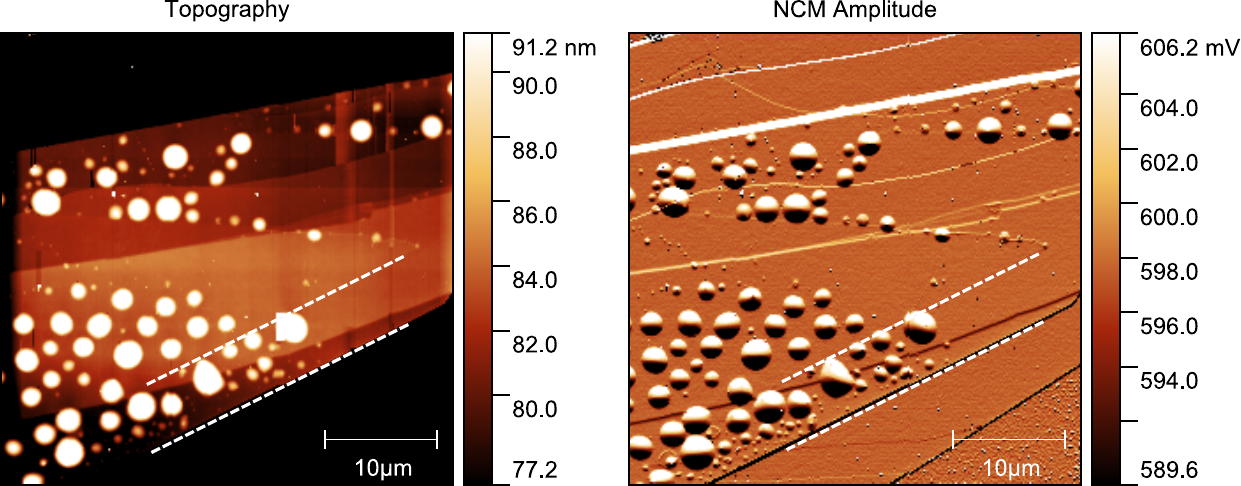}
    \caption{\textbf{AFM of Stack 1} Topography (left) and Amplitude (right) from a tapping mode AFM measurement of the stack described in the main text. The graphene and hBN edges which were identified as zigzag edges with polarized Raman and SHG measurements respectively, and visually aligned during stacking, are indicated by the white dashed lines. These lines are measured to have a $\sim1.2^\circ$ angle between them.
    }
    \label{fig:Stack1AFM}
\end{figure}

 The angle between edges that were identified as crystallographic zigzag edges in optical pre-stack characterization agrees with the twist angle measured in TFM, and falls within the range of twist angles extracted from transport measurements on the two Hall bars. This further validates pre-stack characterization techniques used.

\section{Aligned graphene-hBN Stack 2 Data}
An additional graphene-hBN stack was fabricated using the same optical pre-characterization and TFM post-characterization techniques in the main text. This stack was not encapsulated and fabricated into a device for transport measurements, but still provides further validation for the process flow outlined in the main text. 

The graphene flakes used in this stack is shown in Figure \ref{fig:Stack2Praman}, along with the polarized Raman pre-characterization results. The flake labelled flake 2 was characterized with 2D Raman map, showing where on the flake the D peak signal was observed most prominently. Flakes 1 and 2 share a common vertical edge, and are confirmed to have the same crystal orientation.

\begin{figure}
    \centering
    \includegraphics[width = \linewidth]{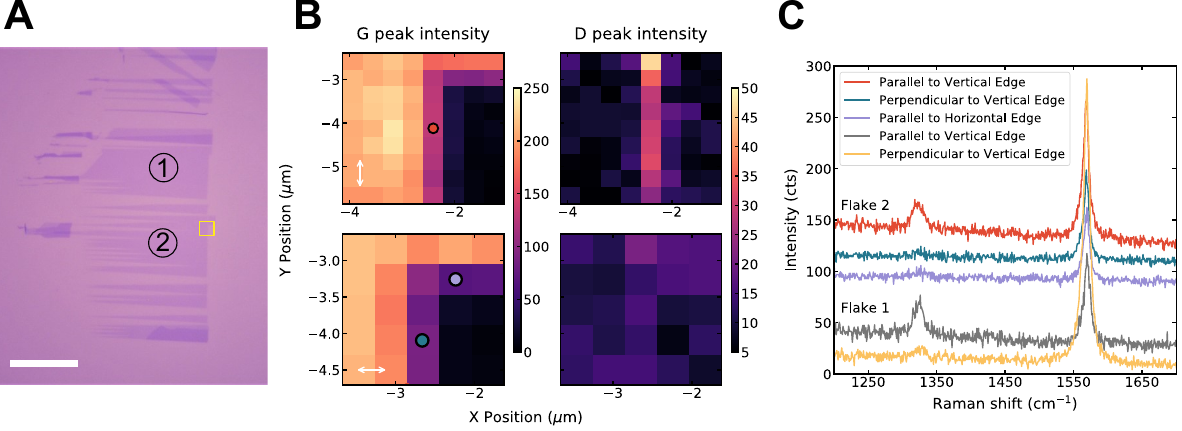}
    \caption{\textbf{Polarized Raman Characterization for Stack 2} 
    \textbf{A)} Optical micrograph of two exfoliated flakes, labelled flake 1 and flake 2. Scale bar is 20 $\mu$m.
    \textbf{B)} Raman maps showing G peak (left) and D peak (right) intensities in the region indicated with the yellow outline in A, where flake 2 has a small 90$^\circ$ corner. Two different maps are taken, one with vertically polarized light (top) and one with horizontally polarized light (bottom) indicated by double headed white arrows. 
    \textbf{C)} Raman spectra taken on both flake 1 and flake 2. The spectra shown from flake 2 are pulled from the map in B - the location of each spectra are indicated by correspondingly colored dots. Two spectra taken along flake 1's vertical edge are also shown (for vertically and horizontally polarized light) to confirm that the two flakes share a crystallographic edge. }
    \label{fig:Stack2Praman}
\end{figure}

The hBN flake used in this stack is shown in Figure \ref{fig:Stack2SHG}, along with the SHG characterization results. 

\begin{figure}
    \centering
    \includegraphics[width = 0.8\linewidth]{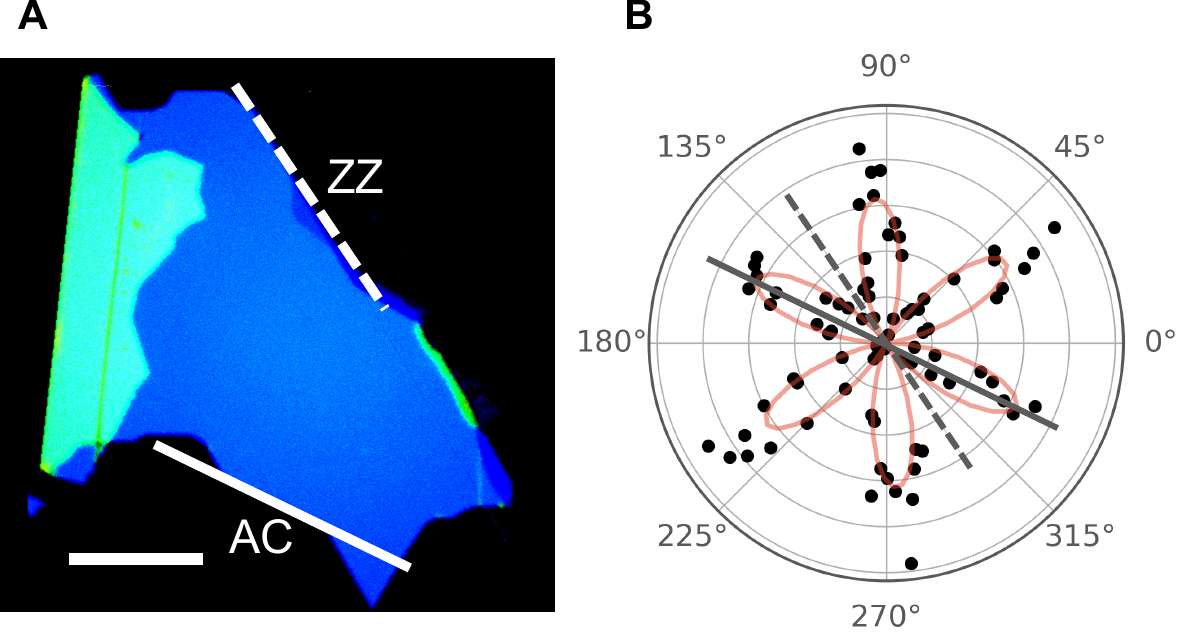}
    \caption{\textbf{SHG Characterization for Stack 2} 
    \textbf{A)} Optical micrograph of hBN flake, with straight edges of interest indicated by white lines. Scale bar is 20 $\mu$m.
    \textbf{B) }Polarization-resolved SHG results taken on this flake. The polarization orientation corresponding to a maximum (node) in the SHG signal intensity are indicated by a solid (dashed) grey line, and are used to label edges in A as armchair (zigzag). 
    }
    \label{fig:Stack2SHG}
\end{figure}

The graphene flakes 1 and 2 are used to create two separate graphene-hBN heterostructures on the same hBN flake using standard dry transfer techniques. First, the hBN flake brought into contact with flake 1, with an armchair edge of flake 1 carefully aligned with an armchair edge of the hBN flake. The hBN flake is then lifted away, picking up graphene flake 1 via van der Waals attraction, forming an aligned graphene-hBN heterostructure and leaving the remaining graphene flakes on the substrate. The substrate is then rotated 30$^\circ$ and a separate region of the hBN flake is brought into contact with flake 2, now with a zigzag edge of the hBN flake aligned with an armchair edge of flake 2, for deliberate 30$^\circ$ misalignement. The Poly(Bisphenol A carbonate) film with the graphene/hBN stack on it is then carefully removed from the PDMS gel/glass slide stamp and placed on top of a SiO$_2$/Si chip to create a open-faced graphene/hBN sample. The chip is then heated to 180 $^{\circ}$C for 2 minutes, during which the polymer film flattens and adheres to the SiO$_2$/Si substrate.

The two graphene-hBN heterostructures are then characterized with TFM (Figure \ref{fig:Stack2TFM}). The deliberately misaligned graphene-hBN heterostructure showed no moir\'e structure, as expected, while the aligned graphene-hBN heterostructure showed a clear moire, validating our armchair/zigzag assignment of graphene and hBN edges. 

\begin{figure}
    \centering
    \includegraphics[width = 0.8\linewidth]{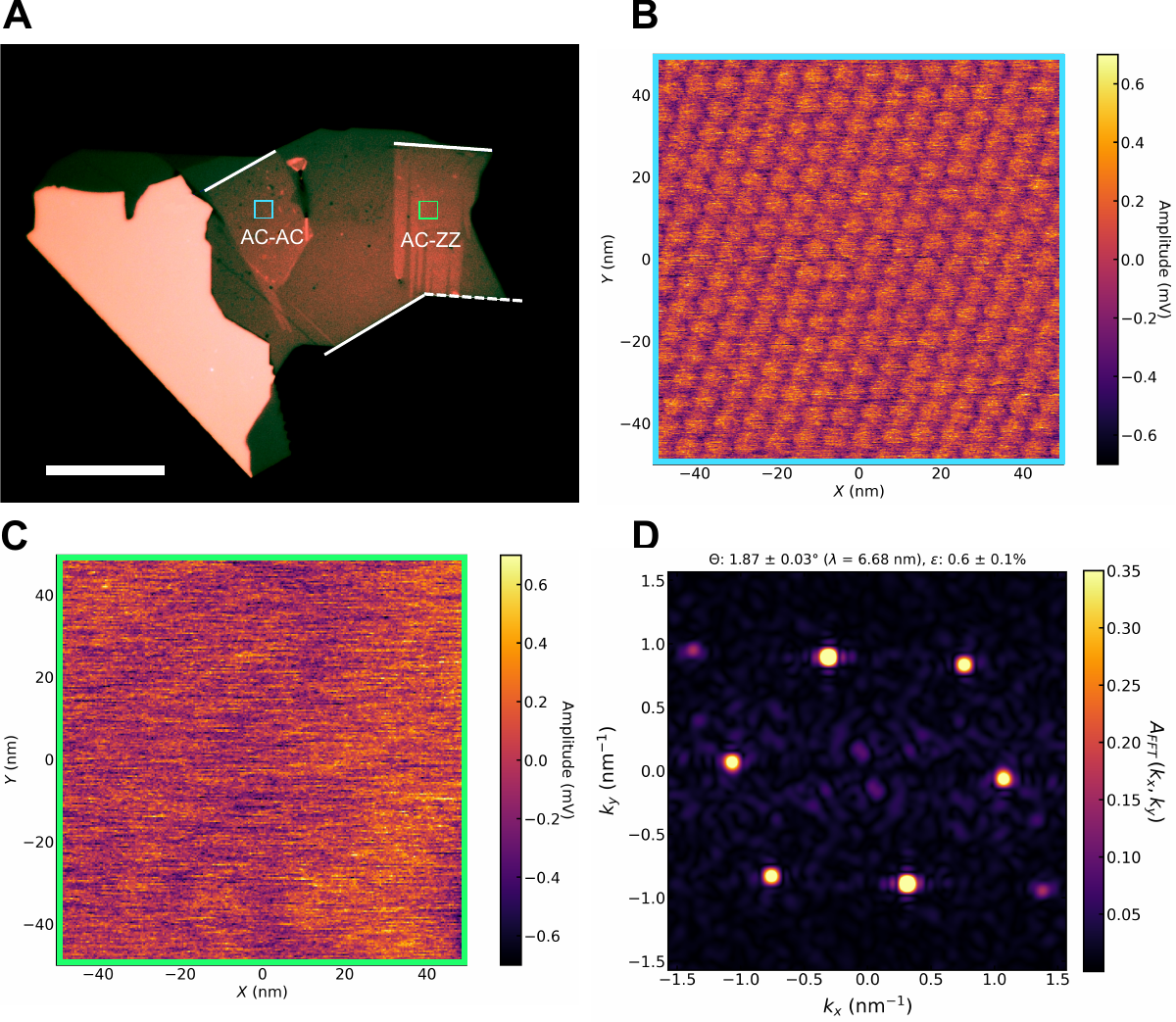}
    \caption{\textbf{TFM Post-Stacking Characterization for Stack 2} 
    \textbf{A)} Optical micrograph of stack 2 on a polymer stamp. Two graphene flakes are visible on the hBN - the armchair edge of the left (right) flake was aligned with an armchair (zigzag) edge of the hBN to form a deliberately aligned (misaligned) heterostructure. Armchair (zigzag) edges used for alignment are indicated by solid (dashed) white lines. 
    \textbf{B)} TFM scan of aligned graphene-hBN heterostructure. Rough location of the 100 nm x 100 nm scan indicated by blue box in A. A clear moir\'e is present.
    \textbf{C)} TFM scan of misaligned graphene-hBN heterostructure. Rough location of the 100 nm x 100 nm scan indicated by green box in A. No moir\'e is visible on this scale.
    \textbf{D)} FFT of TFM image in B. Locations of peaks are use to fit a simple model using twist angle $\theta$ and uniaxial heterostrain $\epsilon$ as fit parameters. Fit results are indicated by blue dots. 
    }
    \label{fig:Stack2TFM}
\end{figure}

We take an FFT of the TFM image and identify the position of six peaks corresponding to the moir\'e reciprocal lattice vectors. These can be fit to a simple model (described below) to determine twist angle and uniaxial heterostrain in the graphene-hBN stack. Like the stack described in the main text, this is a larger twist angle than we'd expect, given that the intended twist angle was zero degrees. However in this case the twist angle error can be in part attributed the fact that the crystal facets that were aligned are relatively short (under 10 microns),  so angular misalignment that would be more obvious with longer straight edges were missed when the facets were aligned by eye during stacking. This source of error can be mitigated in the future by using a high magnification, high resolution optical image to carefully measure graphene and hBN edge orientations with respect to a common long, flat surface before stacking, to reduce human error in alignment during stacking.  

\section{FFT Peak Fitting}
To extract moir\'e parameters from FFT peak positions, we calculate the expected FFT peak positions (corresponding to the two moir\'e reciprocal lattice vectors and their sum) as a function of some initial guess of the following parameters: global crystal orientation ($\Psi$), relative twist ($\theta$), and uniaxial heterostrain magnitude ($\epsilon$) and direction ($\phi$). We then use a least squares fitting method, which steps through the parameter space until the calculated FFT peak positions match the peaks shown in Figure \ref{fig:TFM}C and Supplementary Figure \ref{fig:Stack2TFM}C, to determine our heterostructure parameters. 

To calculate expected FFT peak positions, we start with the unit cell lattice vectors for graphene and hBN: $\vec{a_1} = (a, 0)$ and $\vec{a_2} = (-a/2, \sqrt{3}a/2)$, where $a$ is the the lattice constant 2.46 \r{A} for graphene and 2.50 \r{A} for hBN. The hBN lattice vectors are then rotated by the global angle $\Psi$, and the graphene lattice vectors are rotated by both $\Psi$ and $\theta$ as well as 'strained' by some magnitude $\epsilon$ along some direction $\phi$. These transformations are performed via matrix multiplication with the matrices below.

Rotation matrix for both rotation by arbitrary angle $\alpha$:
\begin{equation}
\label{eq:rot}
    R(\alpha) = 
    \begin{pmatrix}
    \cos(\alpha) & -\sin(\alpha) \\
    \sin(\alpha) & \cos(\alpha)
    \end{pmatrix}
\end{equation}

Uniaxial heterostrain matrix:

\begin{equation}
    S(\epsilon, \psi) = I + R(\psi)^T
    \begin{pmatrix}
        \epsilon & 0 \\
        0 & -0.16\epsilon
    \end{pmatrix}
    R(\psi)
\end{equation}

$S(\epsilon, \psi)$ stretches the lattice by a factor $\epsilon$ along direction $\psi$, and contracts it by $0.16\epsilon$ perpendicular to $\psi$, where 0.16 is the estimated Poisson ratio of graphene. Note - experimental measurements of the Poisson ratio of graphene have measured values ranging from 0.13 to 0.19\cite{blakslee_elastic_2003, seldin_elastic_2003, bosak_elasticity_2007,li_probing_2013}.

After transforming the graphene and hBN lattice vectors, the reciprocal lattice vectors of the individual lattices are calculated using the simple expressions
\begin{equation}
    \vec{b}_1 = 2\pi \frac{R(\frac{\pi}{2})\vec{a}_2}{\vec{a}_1 \cdot R(\frac{\pi}{2})\vec{a}_2}
\end{equation}
\begin{equation}
    \vec{b}_2 = 2\pi \frac{R(\frac{\pi}{2})\vec{a}_1}{\vec{a}_2 \cdot R(\frac{\pi}{2})\vec{a}_2}
\end{equation}

The moir\'e reciprocal lattice vectors are then calculated using the difference between the graphene and hBN reciprocal lattice vectors:

\begin{equation}
    \vec{b}_{M1} = \vec{b}_{g1} - \vec{b}_{hBN1}
\end{equation}
\begin{equation}
    \vec{b}_{M2} = \vec{b}_{g2} - \vec{b}_{hBN2}
\end{equation}

The six FFT peak positions are thus given by $\pm \vec{b}_{M1}$, $\pm \vec{b}_{M2}$, and $\pm (\vec{b}_{M1} - \vec{b}_{M2})$.

After making some initial guess of fit parameters $\Phi$, $\theta$, $\epsilon$, and $\psi$, a least squares fit searches the parameter space until the calculated moir\'e reciprocal lattice vectors match the peak positions in the FFT of the TFM scan. The reported uncertainty values come from the Jacobian matrix reported by the least squares fit algorithm from the scipy.optimize library.

\section{Polarized Raman Additional Information}

\begin{figure}
    \centering
    \includegraphics[width = 0.8\linewidth]{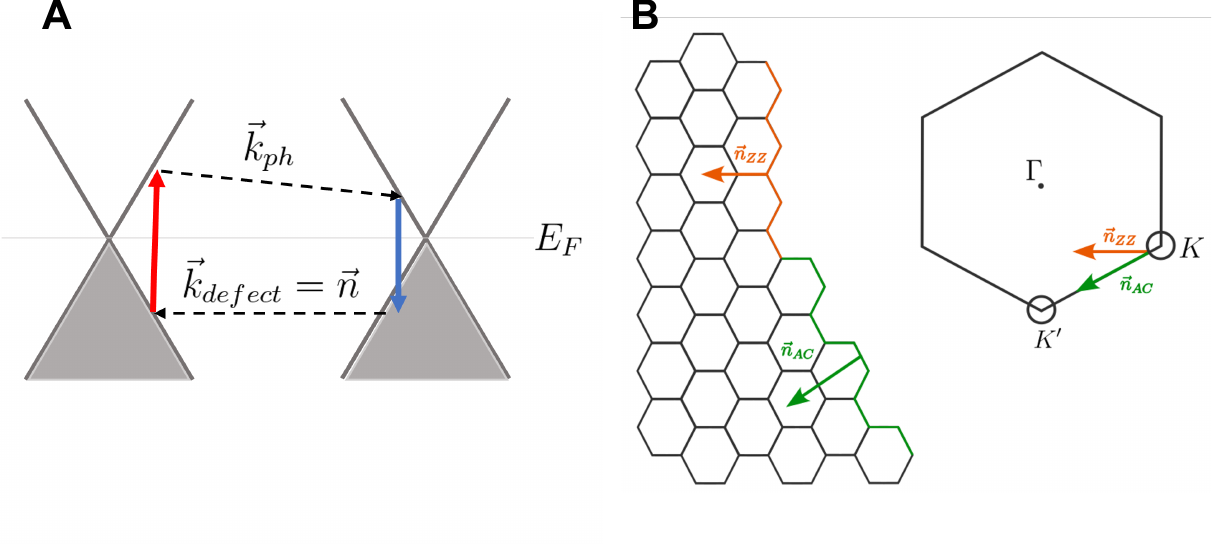}
    \caption{\textbf{Edge-induced Double Resonant Raman Scattering in Graphene.} 
    A) Schematic of the double resonant Raman scattering process behind the D peak in graphene's Raman spectrum. 
    Red (blue) arrow indicates carrier excitation (relaxtion) via photon absorption (emission).
    Dashed lines indicate scattering of the excited carrier with a phonon ($\vec{k}_{ph}$) or a defect/edge ($\vec{k}_{defect} = \vec{n}$).
    B) Crystal structure of graphene and momentum transfer direction associated with scattering off a zigzag (orange) or armchair (green) edge. 
    Edge normal vectors are super-imposed on the graphene Brillioun zone, showing that armchair edges can facilitate intervalley scattering while zigzag edges cannot.
    }
    \label{fig:ZZAC}
\end{figure}

Polarized Raman characterization results for a number of additional graphene flakes are shown here. The results shown in Figure \ref{fig:SuppRamanSuccess} enable unambiguous assignment of edges to armchair and zigzag edge terminations, while the results shown in Figure \ref{fig:SuppRamanUnsuccess} are less clear. 

\begin{figure}
    \centering
    \includegraphics[width=350pt]{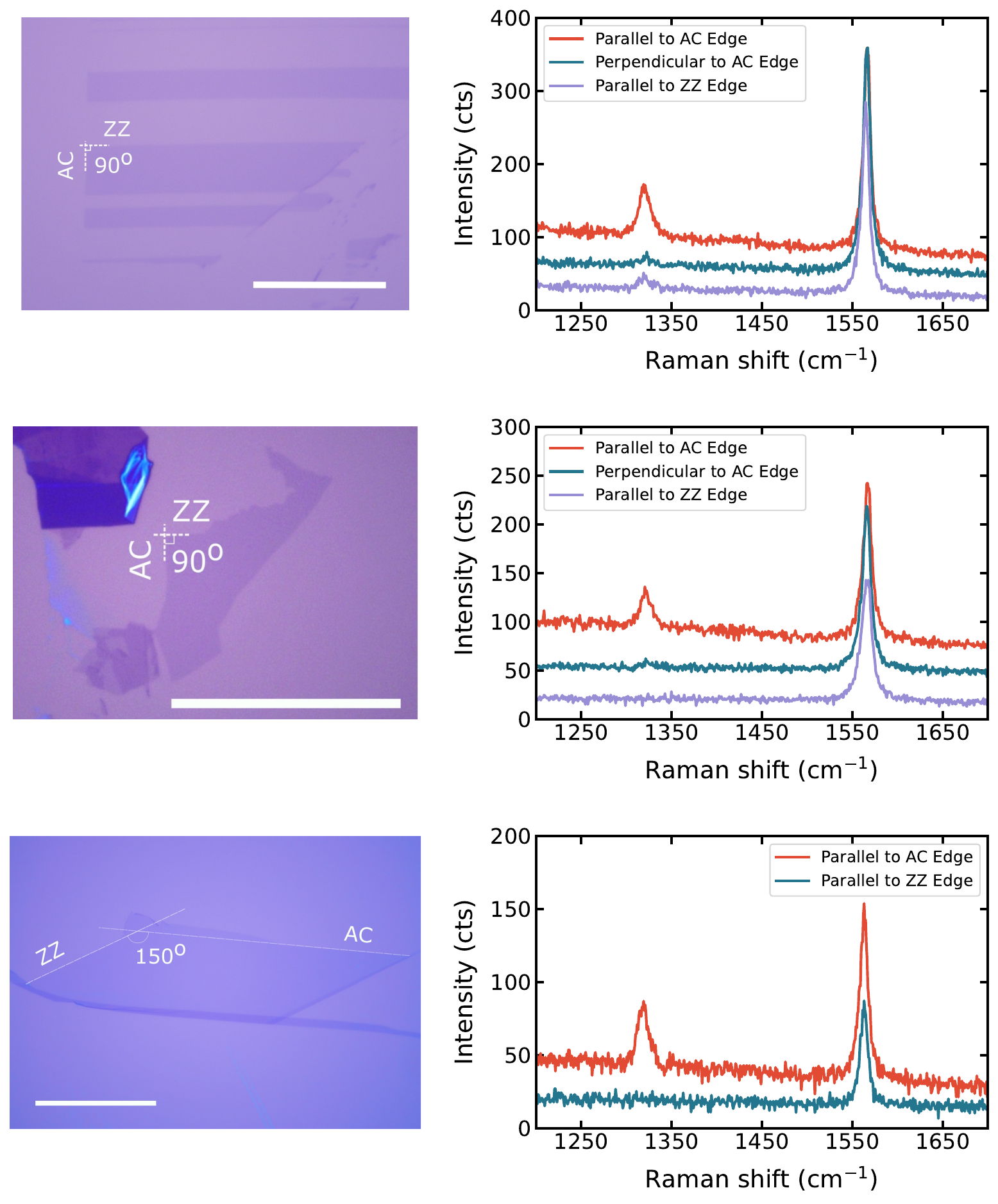}
    \caption{\textbf{Additional Examples of Successful Polarized Raman Characterization} 
    Three different monolayer graphene flakes, each with at least two straight edges separated by an odd integer multiple of 30$^\circ$. For each of these flakes, one of the identified straight edges shows a prominent D peak signal for laser polarization oriented along the edge, while the other does not.  This allows for clear identification of armchair and zigzag edges. In two out of the three flakes, the D peak signal disappears when the laser polarization is rotated perpendicular to the edge, indicating a low disorder edge. 
    }
    \label{fig:SuppRamanSuccess}
\end{figure}

\begin{figure}
    \centering
    \includegraphics[width=350pt]{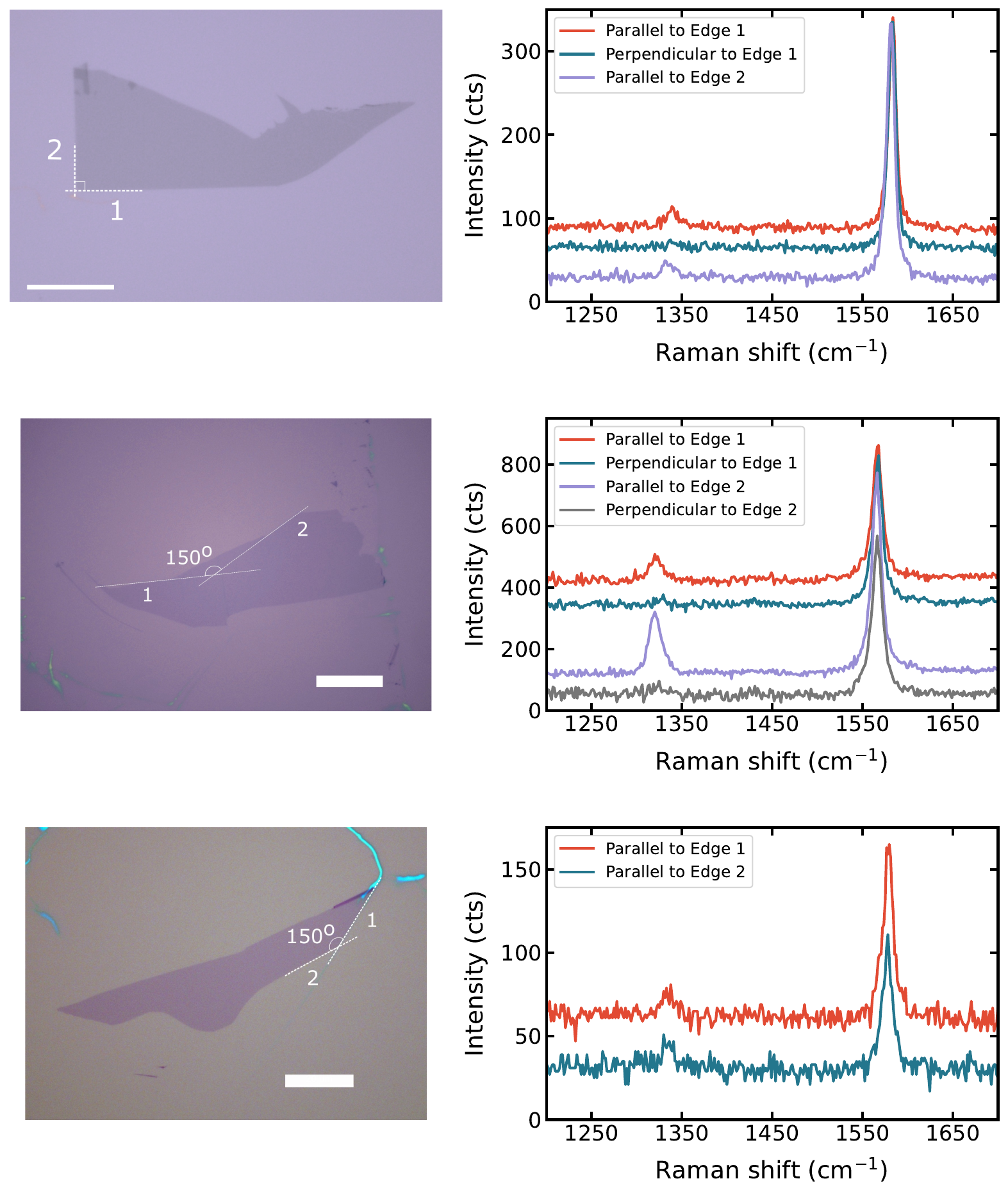}
    \caption{\textbf{Examples of Ambiguous Polarized Raman Characterization} 
    Three different monolayer graphene flakes, also with at least two straight edges separated by an odd integer multiple of $30^\circ$. Though the edges would appear to be crystallographic, Raman characterization of the flakes yielded ambiguous results, with both of the indicated edges showing some indication of a D peak signal. This is likely be the result of zigzag edges that are highly disordered and contain armchair segments that yield a D peak, but it is also possible that that the apparently straight edges indicated here are not crystallographic at all. 
    }
    \label{fig:SuppRamanUnsuccess}
\end{figure}

\section{Second Harmonic Generation Additional Data}

Polarization resolved second harmonic generation measurements were performed on several hBN flakes, the results of which are shown in Figure \ref{fig:SuppSHG}. 
\begin{figure}
    \centering
    \includegraphics[width=325pt]{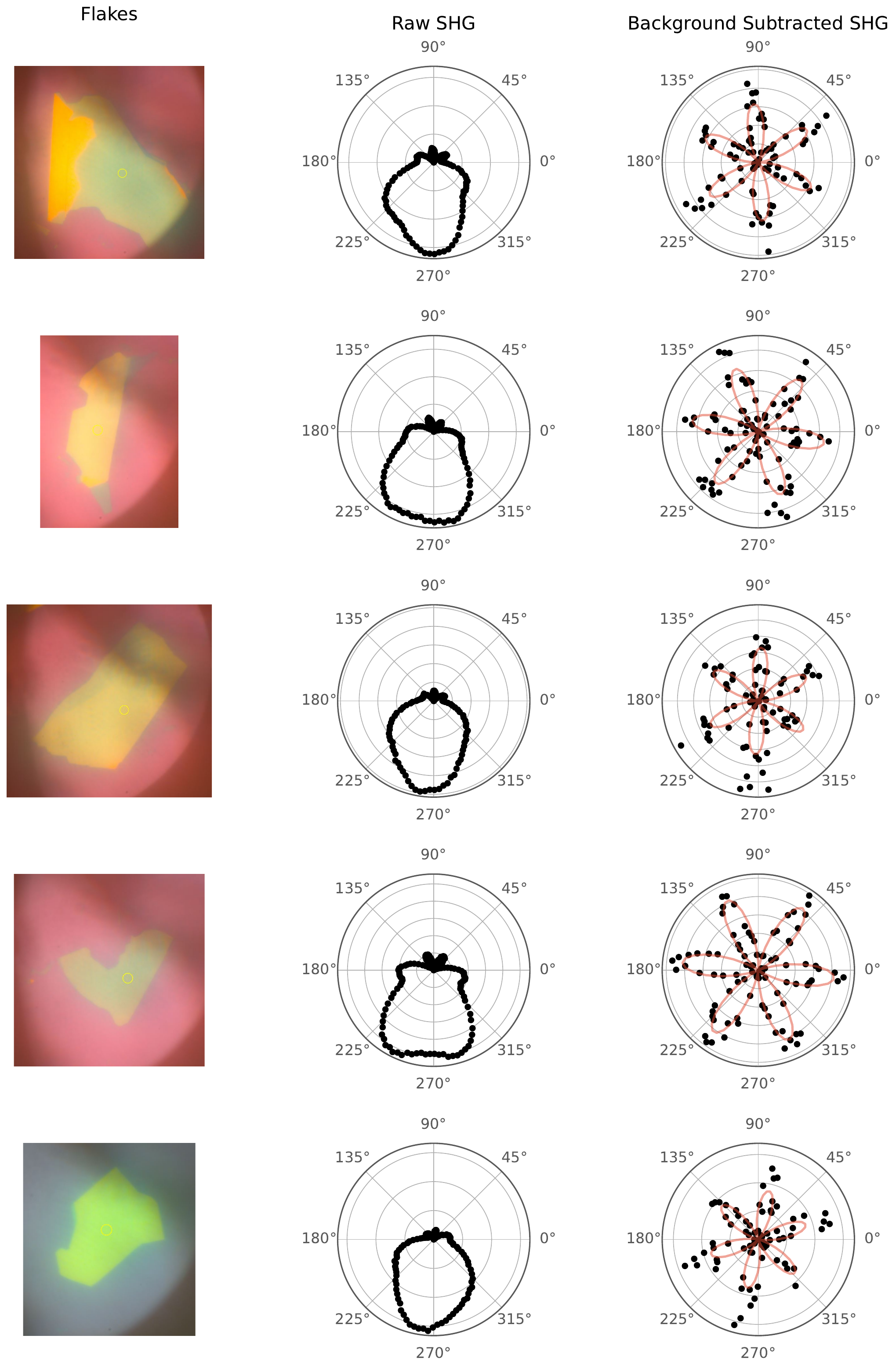}
    \caption{\textbf{Additional SHG Data} 
    SHG data taken on five hBN flakes. Left - optical image of hBN flake within the SHG measurement setup. Middle - raw SHG data for each flake, which includes SHG signal from the flake as well as an SHG signal from the measurement setup half waveplate. Right - Background subtracted SHG data with fit to $I = I_0 \cos^2(3\theta)$.}
    \label{fig:SuppSHG}
\end{figure}

\end{suppinfo}

\bibliography{citations, references}

\providecommand{\latin}[1]{#1}
\makeatletter
\providecommand{\doi}
  {\begingroup\let\do\@makeother\dospecials
  \catcode`\{=1 \catcode`\}=2 \doi@aux}
\providecommand{\doi@aux}[1]{\endgroup\texttt{#1}}
\makeatother
\providecommand*\mcitethebibliography{\thebibliography}
\csname @ifundefined\endcsname{endmcitethebibliography}
  {\let\endmcitethebibliography\endthebibliography}{}
\begin{mcitethebibliography}{53}
\providecommand*\natexlab[1]{#1}
\providecommand*\mciteSetBstSublistMode[1]{}
\providecommand*\mciteSetBstMaxWidthForm[2]{}
\providecommand*\mciteBstWouldAddEndPuncttrue
  {\def\EndOfBibitem{\unskip.}}
\providecommand*\mciteBstWouldAddEndPunctfalse
  {\let\EndOfBibitem\relax}
\providecommand*\mciteSetBstMidEndSepPunct[3]{}
\providecommand*\mciteSetBstSublistLabelBeginEnd[3]{}
\providecommand*\EndOfBibitem{}
\mciteSetBstSublistMode{f}
\mciteSetBstMaxWidthForm{subitem}{(\alph{mcitesubitemcount})}
\mciteSetBstSublistLabelBeginEnd
  {\mcitemaxwidthsubitemform\space}
  {\relax}
  {\relax}

\bibitem[Kim \latin{et~al.}(2016)Kim, Yankowitz, Fallahazad, Kang, Movva,
  Huang, Larentis, Corbet, Taniguchi, Watanabe, and {others}]{kim_van_2016}
Kim,~K.; Yankowitz,~M.; Fallahazad,~B.; Kang,~S.; Movva,~H.~C.; Huang,~S.;
  Larentis,~S.; Corbet,~C.~M.; Taniguchi,~T.; Watanabe,~K.; {others}, van der
  {Waals} heterostructures with high accuracy rotational alignment. \emph{Nano
  letters} \textbf{2016}, \emph{16}, 1989--1995, Publisher: ACS
  Publications\relax
\mciteBstWouldAddEndPuncttrue
\mciteSetBstMidEndSepPunct{\mcitedefaultmidpunct}
{\mcitedefaultendpunct}{\mcitedefaultseppunct}\relax
\EndOfBibitem
\bibitem[Cao \latin{et~al.}(2016)Cao, Luo, Fatemi, Fang, Sanchez-Yamagishi,
  Watanabe, Taniguchi, Kaxiras, and
  Jarillo-Herrero]{cao_superlattice-induced_2016}
Cao,~Y.; Luo,~J.; Fatemi,~V.; Fang,~S.; Sanchez-Yamagishi,~J.; Watanabe,~K.;
  Taniguchi,~T.; Kaxiras,~E.; Jarillo-Herrero,~P. Superlattice-induced
  insulating states and valley-protected orbits in twisted bilayer graphene.
  \emph{Physical review letters} \textbf{2016}, \emph{117}, 116804, Publisher:
  APS\relax
\mciteBstWouldAddEndPuncttrue
\mciteSetBstMidEndSepPunct{\mcitedefaultmidpunct}
{\mcitedefaultendpunct}{\mcitedefaultseppunct}\relax
\EndOfBibitem
\bibitem[Cao \latin{et~al.}(2018)Cao, Fatemi, Demir, Fang, Tomarken, Luo,
  Sanchez-Yamagishi, Watanabe, Taniguchi, Kaxiras, Ashoori, and
  Jarillo-Herrero]{cao_correlated_2018}
Cao,~Y.; Fatemi,~V.; Demir,~A.; Fang,~S.; Tomarken,~S.~L.; Luo,~J.~Y.;
  Sanchez-Yamagishi,~J.~D.; Watanabe,~K.; Taniguchi,~T.; Kaxiras,~E.;
  Ashoori,~R.~C.; Jarillo-Herrero,~P. Correlated insulator behaviour at
  half-filling in magic-angle graphene superlattices. \emph{Nature}
  \textbf{2018}, \emph{556}, 80--84, Number: 7699 Publisher: Nature Publishing
  Group\relax
\mciteBstWouldAddEndPuncttrue
\mciteSetBstMidEndSepPunct{\mcitedefaultmidpunct}
{\mcitedefaultendpunct}{\mcitedefaultseppunct}\relax
\EndOfBibitem
\bibitem[Cao \latin{et~al.}(2018)Cao, Fatemi, Fang, Watanabe, Taniguchi,
  Kaxiras, and Jarillo-Herrero]{cao_unconventional_2018}
Cao,~Y.; Fatemi,~V.; Fang,~S.; Watanabe,~K.; Taniguchi,~T.; Kaxiras,~E.;
  Jarillo-Herrero,~P. Unconventional superconductivity in magic-angle graphene
  superlattices. \emph{Nature} \textbf{2018}, \emph{556}, 43--50, Number: 7699
  Publisher: Nature Publishing Group\relax
\mciteBstWouldAddEndPuncttrue
\mciteSetBstMidEndSepPunct{\mcitedefaultmidpunct}
{\mcitedefaultendpunct}{\mcitedefaultseppunct}\relax
\EndOfBibitem
\bibitem[Yankowitz \latin{et~al.}(2019)Yankowitz, Chen, Polshyn, Zhang,
  Watanabe, Taniguchi, Graf, Young, and Dean]{yankowitz_tuning_2019}
Yankowitz,~M.; Chen,~S.; Polshyn,~H.; Zhang,~Y.; Watanabe,~K.; Taniguchi,~T.;
  Graf,~D.; Young,~A.~F.; Dean,~C.~R. Tuning superconductivity in twisted
  bilayer graphene. \emph{Science} \textbf{2019}, \emph{363}, 1059--1064,
  Publisher: American Association for the Advancement of Science\relax
\mciteBstWouldAddEndPuncttrue
\mciteSetBstMidEndSepPunct{\mcitedefaultmidpunct}
{\mcitedefaultendpunct}{\mcitedefaultseppunct}\relax
\EndOfBibitem
\bibitem[Lu \latin{et~al.}(2019)Lu, Stepanov, Yang, Xie, Aamir, Das, Urgell,
  Watanabe, Taniguchi, Zhang, Bachtold, MacDonald, and
  Efetov]{lu_superconductors_2019}
Lu,~X.; Stepanov,~P.; Yang,~W.; Xie,~M.; Aamir,~M.~A.; Das,~I.; Urgell,~C.;
  Watanabe,~K.; Taniguchi,~T.; Zhang,~G.; Bachtold,~A.; MacDonald,~A.~H.;
  Efetov,~D.~K. Superconductors, orbital magnets and correlated states in
  magic-angle bilayer graphene. \emph{Nature} \textbf{2019}, \emph{574},
  653--657, Number: 7780 Publisher: Nature Publishing Group\relax
\mciteBstWouldAddEndPuncttrue
\mciteSetBstMidEndSepPunct{\mcitedefaultmidpunct}
{\mcitedefaultendpunct}{\mcitedefaultseppunct}\relax
\EndOfBibitem
\bibitem[Stepanov \latin{et~al.}(2020)Stepanov, Das, Lu, Fahimniya, Watanabe,
  Taniguchi, Koppens, Lischner, Levitov, and Efetov]{stepanov_untying_2020}
Stepanov,~P.; Das,~I.; Lu,~X.; Fahimniya,~A.; Watanabe,~K.; Taniguchi,~T.;
  Koppens,~F. H.~L.; Lischner,~J.; Levitov,~L.; Efetov,~D.~K. Untying the
  insulating and superconducting orders in magic-angle graphene. \emph{Nature}
  \textbf{2020}, \emph{583}, 375--378, Number: 7816 Publisher: Nature
  Publishing Group\relax
\mciteBstWouldAddEndPuncttrue
\mciteSetBstMidEndSepPunct{\mcitedefaultmidpunct}
{\mcitedefaultendpunct}{\mcitedefaultseppunct}\relax
\EndOfBibitem
\bibitem[Moon and Koshino(2014)Moon, and Koshino]{moon_electronic_2014}
Moon,~P.; Koshino,~M. Electronic properties of graphene/hexagonal-boron-nitride
  moir{\textbackslash}'e superlattice. \emph{Physical Review B} \textbf{2014},
  \emph{90}, 155406, Publisher: American Physical Society\relax
\mciteBstWouldAddEndPuncttrue
\mciteSetBstMidEndSepPunct{\mcitedefaultmidpunct}
{\mcitedefaultendpunct}{\mcitedefaultseppunct}\relax
\EndOfBibitem
\bibitem[Yankowitz \latin{et~al.}(2012)Yankowitz, Xue, Cormode,
  Sanchez-Yamagishi, Watanabe, Taniguchi, Jarillo-Herrero, Jacquod, and
  LeRoy]{yankowitz_emergence_2012}
Yankowitz,~M.; Xue,~J.; Cormode,~D.; Sanchez-Yamagishi,~J.~D.; Watanabe,~K.;
  Taniguchi,~T.; Jarillo-Herrero,~P.; Jacquod,~P.; LeRoy,~B.~J. Emergence of
  superlattice {Dirac} points in graphene on hexagonal boron nitride.
  \emph{Nature Physics} \textbf{2012}, \emph{8}, 382--386, Number: 5 Publisher:
  Nature Publishing Group\relax
\mciteBstWouldAddEndPuncttrue
\mciteSetBstMidEndSepPunct{\mcitedefaultmidpunct}
{\mcitedefaultendpunct}{\mcitedefaultseppunct}\relax
\EndOfBibitem
\bibitem[Lee \latin{et~al.}(2016)Lee, Wallbank, Gallagher, Watanabe, Taniguchi,
  Fal’ko, and Goldhaber-Gordon]{lee_ballistic_2016}
Lee,~M.; Wallbank,~J.~R.; Gallagher,~P.; Watanabe,~K.; Taniguchi,~T.;
  Fal’ko,~V.~I.; Goldhaber-Gordon,~D. Ballistic miniband conduction in a
  graphene superlattice. \emph{Science} \textbf{2016}, \emph{353}, 1526--1529,
  Publisher: American Association for the Advancement of Science\relax
\mciteBstWouldAddEndPuncttrue
\mciteSetBstMidEndSepPunct{\mcitedefaultmidpunct}
{\mcitedefaultendpunct}{\mcitedefaultseppunct}\relax
\EndOfBibitem
\bibitem[Ponomarenko \latin{et~al.}(2013)Ponomarenko, Gorbachev, Yu, Elias,
  Jalil, Patel, Mishchenko, Mayorov, Woods, Wallbank, Mucha-Kruczynski, Piot,
  Potemski, Grigorieva, Novoselov, Guinea, Falâ€™ko, and
  Geim]{ponomarenko_cloning_2013}
Ponomarenko,~L.~A. \latin{et~al.}  Cloning of {Dirac} fermions in graphene
  superlattices. \emph{Nature} \textbf{2013}, \emph{497}, 594--597\relax
\mciteBstWouldAddEndPuncttrue
\mciteSetBstMidEndSepPunct{\mcitedefaultmidpunct}
{\mcitedefaultendpunct}{\mcitedefaultseppunct}\relax
\EndOfBibitem
\bibitem[Dean \latin{et~al.}(2013)Dean, Wang, Maher, Forsythe, Ghahari, Gao,
  Katoch, Ishigami, Moon, Koshino, Taniguchi, Watanabe, Shepard, Hone, and
  Kim]{dean_hofstadters_2013}
Dean,~C.~R.; Wang,~L.; Maher,~P.; Forsythe,~C.; Ghahari,~F.; Gao,~Y.;
  Katoch,~J.; Ishigami,~M.; Moon,~P.; Koshino,~M.; Taniguchi,~T.; Watanabe,~K.;
  Shepard,~K.~L.; Hone,~J.; Kim,~P. Hofstadter’s butterfly and the fractal
  quantum {Hall} effect in moiré superlattices. \emph{Nature} \textbf{2013},
  \emph{497}, 598--602, Number: 7451 Publisher: Nature Publishing Group\relax
\mciteBstWouldAddEndPuncttrue
\mciteSetBstMidEndSepPunct{\mcitedefaultmidpunct}
{\mcitedefaultendpunct}{\mcitedefaultseppunct}\relax
\EndOfBibitem
\bibitem[Hunt \latin{et~al.}(2013)Hunt, Sanchez-Yamagishi, Young, Yankowitz,
  LeRoy, Watanabe, Taniguchi, Moon, Koshino, Jarillo-Herrero, and
  Ashoori]{hunt_massive_2013}
Hunt,~B.; Sanchez-Yamagishi,~J.~D.; Young,~A.~F.; Yankowitz,~M.; LeRoy,~B.~J.;
  Watanabe,~K.; Taniguchi,~T.; Moon,~P.; Koshino,~M.; Jarillo-Herrero,~P.;
  Ashoori,~R.~C. Massive {Dirac} {Fermions} and {Hofstadter} {Butterfly} in a
  van der {Waals} {Heterostructure}. \emph{Science} \textbf{2013}, \emph{340},
  1427--1430, Publisher: American Association for the Advancement of
  Science\relax
\mciteBstWouldAddEndPuncttrue
\mciteSetBstMidEndSepPunct{\mcitedefaultmidpunct}
{\mcitedefaultendpunct}{\mcitedefaultseppunct}\relax
\EndOfBibitem
\bibitem[Yu \latin{et~al.}(2014)Yu, Gorbachev, Tu, Kretinin, Cao, Jalil,
  Withers, Ponomarenko, Piot, Potemski, and {others}]{yu_hierarchy_2014}
Yu,~G.; Gorbachev,~R.; Tu,~J.; Kretinin,~A.; Cao,~Y.; Jalil,~R.; Withers,~F.;
  Ponomarenko,~L.; Piot,~B.; Potemski,~M.; {others}, Hierarchy of {Hofstadter}
  states and replica quantum {Hall} ferromagnetism in graphene superlattices.
  \emph{Nature physics} \textbf{2014}, \emph{10}, 525--529, Publisher: Nature
  Publishing Group\relax
\mciteBstWouldAddEndPuncttrue
\mciteSetBstMidEndSepPunct{\mcitedefaultmidpunct}
{\mcitedefaultendpunct}{\mcitedefaultseppunct}\relax
\EndOfBibitem
\bibitem[Yao \latin{et~al.}(2021)Yao, Finney, Zhang, Moore, Xian,
  Tancogne-Dejean, Liu, Ardelean, Xu, Halbertal, Watanabe, Taniguchi, Ochoa,
  Asenjo-Garcia, Zhu, Basov, Rubio, Dean, Hone, and Schuck]{yao_enhanced_2021}
Yao,~K. \latin{et~al.}  Enhanced tunable second harmonic generation from
  twistable interfaces and vertical superlattices in boron nitride
  homostructures. \emph{Science Advances} \textbf{2021}, \emph{7}, eabe8691,
  \_eprint: https://www.science.org/doi/pdf/10.1126/sciadv.abe8691\relax
\mciteBstWouldAddEndPuncttrue
\mciteSetBstMidEndSepPunct{\mcitedefaultmidpunct}
{\mcitedefaultendpunct}{\mcitedefaultseppunct}\relax
\EndOfBibitem
\bibitem[Jat \latin{et~al.}(2024)Jat, Tiwari, Bajaj, Shitut, Mandal, Watanabe,
  Taniguchi, Krishnamurthy, Jain, and Bid]{jat_higher_2024}
Jat,~M.~K.; Tiwari,~P.; Bajaj,~R.; Shitut,~I.; Mandal,~S.; Watanabe,~K.;
  Taniguchi,~T.; Krishnamurthy,~H.~R.; Jain,~M.; Bid,~A. Higher order gaps in
  the renormalized band structure of doubly aligned {hBN}/bilayer graphene
  moiré superlattice. \emph{Nature Communications} \textbf{2024}, \emph{15},
  2335, Publisher: Nature Publishing Group\relax
\mciteBstWouldAddEndPuncttrue
\mciteSetBstMidEndSepPunct{\mcitedefaultmidpunct}
{\mcitedefaultendpunct}{\mcitedefaultseppunct}\relax
\EndOfBibitem
\bibitem[Sharpe \latin{et~al.}(2019)Sharpe, Fox, Barnard, Finney, Watanabe,
  Taniguchi, Kastner, and Goldhaber-Gordon]{sharpe_emergent_2019}
Sharpe,~A.~L.; Fox,~E.~J.; Barnard,~A.~W.; Finney,~J.; Watanabe,~K.;
  Taniguchi,~T.; Kastner,~M.~A.; Goldhaber-Gordon,~D. Emergent ferromagnetism
  near three-quarters filling in twisted bilayer graphene. \emph{Science}
  \textbf{2019}, \emph{365}, 605--608\relax
\mciteBstWouldAddEndPuncttrue
\mciteSetBstMidEndSepPunct{\mcitedefaultmidpunct}
{\mcitedefaultendpunct}{\mcitedefaultseppunct}\relax
\EndOfBibitem
\bibitem[Serlin \latin{et~al.}(2020)Serlin, Tschirhart, Polshyn, Zhang, Zhu,
  Watanabe, Taniguchi, Balents, and Young]{serlin_intrinsic_2020}
Serlin,~M.; Tschirhart,~C.~L.; Polshyn,~H.; Zhang,~Y.; Zhu,~J.; Watanabe,~K.;
  Taniguchi,~T.; Balents,~L.; Young,~A.~F. Intrinsic quantized anomalous {Hall}
  effect in a moiré heterostructure. \emph{Science} \textbf{2020}, \emph{367},
  900--903, Publisher: American Association for the Advancement of
  Science\relax
\mciteBstWouldAddEndPuncttrue
\mciteSetBstMidEndSepPunct{\mcitedefaultmidpunct}
{\mcitedefaultendpunct}{\mcitedefaultseppunct}\relax
\EndOfBibitem
\bibitem[Chen \latin{et~al.}(2019)Chen, Jiang, Wu, Lyu, Li, Chittari, Watanabe,
  Taniguchi, Shi, Jung, Zhang, and Wang]{chen_evidence_2019}
Chen,~G.; Jiang,~L.; Wu,~S.; Lyu,~B.; Li,~H.; Chittari,~B.~L.; Watanabe,~K.;
  Taniguchi,~T.; Shi,~Z.; Jung,~J.; Zhang,~Y.; Wang,~F. Evidence of a
  gate-tunable {Mott} insulator in a trilayer graphene moiré superlattice.
  \emph{Nature Physics} \textbf{2019}, \emph{15}, 237--241, Publisher: Nature
  Publishing Group\relax
\mciteBstWouldAddEndPuncttrue
\mciteSetBstMidEndSepPunct{\mcitedefaultmidpunct}
{\mcitedefaultendpunct}{\mcitedefaultseppunct}\relax
\EndOfBibitem
\bibitem[Chen \latin{et~al.}(2019)Chen, Sharpe, Gallagher, Rosen, Fox, Jiang,
  Lyu, Li, Watanabe, Taniguchi, Jung, Shi, Goldhaber-Gordon, Zhang, and
  Wang]{chen_signatures_2019}
Chen,~G.; Sharpe,~A.~L.; Gallagher,~P.; Rosen,~I.~T.; Fox,~E.~J.; Jiang,~L.;
  Lyu,~B.; Li,~H.; Watanabe,~K.; Taniguchi,~T.; Jung,~J.; Shi,~Z.;
  Goldhaber-Gordon,~D.; Zhang,~Y.; Wang,~F. Signatures of tunable
  superconductivity in a trilayer graphene moiré superlattice. \emph{Nature}
  \textbf{2019}, \emph{572}, 215--219, Publisher: Nature Publishing Group\relax
\mciteBstWouldAddEndPuncttrue
\mciteSetBstMidEndSepPunct{\mcitedefaultmidpunct}
{\mcitedefaultendpunct}{\mcitedefaultseppunct}\relax
\EndOfBibitem
\bibitem[Chen \latin{et~al.}(2020)Chen, Sharpe, Fox, Zhang, Wang, Jiang, Lyu,
  Li, Watanabe, Taniguchi, Shi, Senthil, Goldhaber-Gordon, Zhang, and
  Wang]{chen_tunable_2020}
Chen,~G.; Sharpe,~A.~L.; Fox,~E.~J.; Zhang,~Y.-H.; Wang,~S.; Jiang,~L.;
  Lyu,~B.; Li,~H.; Watanabe,~K.; Taniguchi,~T.; Shi,~Z.; Senthil,~T.;
  Goldhaber-Gordon,~D.; Zhang,~Y.; Wang,~F. Tunable correlated {Chern}
  insulator and ferromagnetism in a moiré superlattice. \emph{Nature}
  \textbf{2020}, \emph{579}, 56--61, Publisher: Nature Publishing Group\relax
\mciteBstWouldAddEndPuncttrue
\mciteSetBstMidEndSepPunct{\mcitedefaultmidpunct}
{\mcitedefaultendpunct}{\mcitedefaultseppunct}\relax
\EndOfBibitem
\bibitem[Lu \latin{et~al.}(2024)Lu, Han, Yao, Reddy, Yang, Seo, Watanabe,
  Taniguchi, Fu, and Ju]{lu_fractional_2024}
Lu,~Z.; Han,~T.; Yao,~Y.; Reddy,~A.~P.; Yang,~J.; Seo,~J.; Watanabe,~K.;
  Taniguchi,~T.; Fu,~L.; Ju,~L. Fractional quantum anomalous {Hall} effect in
  multilayer graphene. \emph{Nature} \textbf{2024}, \emph{626}, 759--764,
  Publisher: Nature Publishing Group\relax
\mciteBstWouldAddEndPuncttrue
\mciteSetBstMidEndSepPunct{\mcitedefaultmidpunct}
{\mcitedefaultendpunct}{\mcitedefaultseppunct}\relax
\EndOfBibitem
\bibitem[Zhang \latin{et~al.}(2019)Zhang, Mao, and Senthil]{zhang_twisted_2019}
Zhang,~Y.-H.; Mao,~D.; Senthil,~T. Twisted bilayer graphene aligned with
  hexagonal boron nitride: {Anomalous} {Hall} effect and a lattice model.
  \emph{Phys. Rev. Res.} \textbf{2019}, \emph{1}, 033126, Publisher: American
  Physical Society\relax
\mciteBstWouldAddEndPuncttrue
\mciteSetBstMidEndSepPunct{\mcitedefaultmidpunct}
{\mcitedefaultendpunct}{\mcitedefaultseppunct}\relax
\EndOfBibitem
\bibitem[Bultinck \latin{et~al.}(2020)Bultinck, Chatterjee, and
  Zaletel]{bultinck_mechanism_2020}
Bultinck,~N.; Chatterjee,~S.; Zaletel,~M.~P. Mechanism for {Anomalous} {Hall}
  {Ferromagnetism} in {Twisted} {Bilayer} {Graphene}. \emph{Physical Review
  Letters} \textbf{2020}, \emph{124}, 166601, Publisher: American Physical
  Society\relax
\mciteBstWouldAddEndPuncttrue
\mciteSetBstMidEndSepPunct{\mcitedefaultmidpunct}
{\mcitedefaultendpunct}{\mcitedefaultseppunct}\relax
\EndOfBibitem
\bibitem[Kindermann \latin{et~al.}(2012)Kindermann, Uchoa, and
  Miller]{kindermann_zero-energy_2012}
Kindermann,~M.; Uchoa,~B.; Miller,~D.~L. Zero-energy modes and gate-tunable gap
  in graphene on hexagonal boron nitride. \emph{Physical Review B}
  \textbf{2012}, \emph{86}, 115415, Publisher: American Physical Society\relax
\mciteBstWouldAddEndPuncttrue
\mciteSetBstMidEndSepPunct{\mcitedefaultmidpunct}
{\mcitedefaultendpunct}{\mcitedefaultseppunct}\relax
\EndOfBibitem
\bibitem[Shi \latin{et~al.}(2021)Shi, Zhu, and MacDonald]{shi_moire_2021}
Shi,~J.; Zhu,~J.; MacDonald,~A.~H. Moiré commensurability and the quantum
  anomalous {Hall} effect in twisted bilayer graphene on hexagonal boron
  nitride. \emph{Phys. Rev. B} \textbf{2021}, \emph{103}, 075122, Publisher:
  American Physical Society\relax
\mciteBstWouldAddEndPuncttrue
\mciteSetBstMidEndSepPunct{\mcitedefaultmidpunct}
{\mcitedefaultendpunct}{\mcitedefaultseppunct}\relax
\EndOfBibitem
\bibitem[Cea \latin{et~al.}(2020)Cea, Pantaleón, and Guinea]{cea_band_2020}
Cea,~T.; Pantaleón,~P.~A.; Guinea,~F. Band structure of twisted bilayer
  graphene on hexagonal boron nitride. \emph{Phys. Rev. B} \textbf{2020},
  \emph{102}, 155136, Publisher: American Physical Society\relax
\mciteBstWouldAddEndPuncttrue
\mciteSetBstMidEndSepPunct{\mcitedefaultmidpunct}
{\mcitedefaultendpunct}{\mcitedefaultseppunct}\relax
\EndOfBibitem
\bibitem[Mao and Senthil(2021)Mao, and Senthil]{mao_quasiperiodicity_2021}
Mao,~D.; Senthil,~T. Quasiperiodicity, band topology, and
  moir{\textbackslash}'e graphene. \emph{Physical Review B} \textbf{2021},
  \emph{103}, 115110, Publisher: American Physical Society\relax
\mciteBstWouldAddEndPuncttrue
\mciteSetBstMidEndSepPunct{\mcitedefaultmidpunct}
{\mcitedefaultendpunct}{\mcitedefaultseppunct}\relax
\EndOfBibitem
\bibitem[Lai \latin{et~al.}(2023)Lai, Guerci, Li, Watanabe, Taniguchi, Wilson,
  Pixley, and Andrei]{lai_imaging_2023}
Lai,~X.; Guerci,~D.; Li,~G.; Watanabe,~K.; Taniguchi,~T.; Wilson,~J.;
  Pixley,~J.~H.; Andrei,~E.~Y. Imaging {Self}-aligned {Moir}{\textbackslash}'e
  {Crystals} and {Quasicrystals} in {Magic}-angle {Bilayer} {Graphene} on {hBN}
  {Heterostructures}. 2023; \url{http://arxiv.org/abs/2311.07819},
  arXiv:2311.07819 [cond-mat] version: 1\relax
\mciteBstWouldAddEndPuncttrue
\mciteSetBstMidEndSepPunct{\mcitedefaultmidpunct}
{\mcitedefaultendpunct}{\mcitedefaultseppunct}\relax
\EndOfBibitem
\bibitem[You \latin{et~al.}(2008)You, Ni, Yu, and Shen]{you_edge_2008}
You,~Y.; Ni,~Z.; Yu,~T.; Shen,~Z. Edge chirality determination of graphene by
  {Raman} spectroscopy. \emph{Applied Physics Letters} \textbf{2008},
  \emph{93}, 163112\relax
\mciteBstWouldAddEndPuncttrue
\mciteSetBstMidEndSepPunct{\mcitedefaultmidpunct}
{\mcitedefaultendpunct}{\mcitedefaultseppunct}\relax
\EndOfBibitem
\bibitem[Neubeck \latin{et~al.}(2010)Neubeck, You, Ni, Blake, Shen, Geim, and
  Novoselov]{neubeck_direct_2010}
Neubeck,~S.; You,~Y.~M.; Ni,~Z.~H.; Blake,~P.; Shen,~Z.~X.; Geim,~A.~K.;
  Novoselov,~K.~S. Direct determination of the crystallographic orientation of
  graphene edges by atomic resolution imaging. \emph{Applied Physics Letters}
  \textbf{2010}, \emph{97}, 053110\relax
\mciteBstWouldAddEndPuncttrue
\mciteSetBstMidEndSepPunct{\mcitedefaultmidpunct}
{\mcitedefaultendpunct}{\mcitedefaultseppunct}\relax
\EndOfBibitem
\bibitem[Kawai \latin{et~al.}(2009)Kawai, Okada, Miyamoto, and
  Hiura]{kawai_self-redirection_2009}
Kawai,~T.; Okada,~S.; Miyamoto,~Y.; Hiura,~H. Self-redirection of tearing edges
  in graphene: {Tight}-binding molecular dynamics simulations. \emph{Physical
  Review B} \textbf{2009}, \emph{80}, 033401, Publisher: American Physical
  Society\relax
\mciteBstWouldAddEndPuncttrue
\mciteSetBstMidEndSepPunct{\mcitedefaultmidpunct}
{\mcitedefaultendpunct}{\mcitedefaultseppunct}\relax
\EndOfBibitem
\bibitem[Kim \latin{et~al.}(2012)Kim, Artyukhov, Regan, Liu, Crommie, Yakobson,
  and Zettl]{kim_ripping_2012}
Kim,~K.; Artyukhov,~V.~I.; Regan,~W.; Liu,~Y.; Crommie,~M.~F.; Yakobson,~B.~I.;
  Zettl,~A. Ripping {Graphene}: {Preferred} {Directions}. \emph{Nano Letters}
  \textbf{2012}, \emph{12}, 293--297, Publisher: American Chemical
  Society\relax
\mciteBstWouldAddEndPuncttrue
\mciteSetBstMidEndSepPunct{\mcitedefaultmidpunct}
{\mcitedefaultendpunct}{\mcitedefaultseppunct}\relax
\EndOfBibitem
\bibitem[Canıfmmode
  {\textbackslash}mboxç{\textbackslash}else~ç{\textbackslash}fiado
  \latin{et~al.}(2004)Canıfmmode
  {\textbackslash}mboxç{\textbackslash}else~ç{\textbackslash}fiado, Pimenta,
  Neves, Dantas, and Jorio]{canifmmode_mboxcelse_cfiado_influence_2004}
Canıfmmode
  {\textbackslash}mboxç{\textbackslash}else~ç{\textbackslash}fiado,~L.~G.;
  Pimenta,~M.~A.; Neves,~B. R.~A.; Dantas,~M. S.~S.; Jorio,~A. Influence of the
  {Atomic} {Structure} on the {Raman} {Spectra} of {Graphite} {Edges}.
  \emph{Phys. Rev. Lett.} \textbf{2004}, \emph{93}, 247401, Publisher: American
  Physical Society\relax
\mciteBstWouldAddEndPuncttrue
\mciteSetBstMidEndSepPunct{\mcitedefaultmidpunct}
{\mcitedefaultendpunct}{\mcitedefaultseppunct}\relax
\EndOfBibitem
\bibitem[Casiraghi \latin{et~al.}(2009)Casiraghi, Hartschuh, Qian, Piscanec,
  Georgi, Fasoli, Novoselov, Basko, and Ferrari]{casiraghi_raman_2009}
Casiraghi,~C.; Hartschuh,~A.; Qian,~H.; Piscanec,~S.; Georgi,~C.; Fasoli,~A.;
  Novoselov,~K.~S.; Basko,~D.~M.; Ferrari,~A.~C. Raman {Spectroscopy} of
  {Graphene} {Edges}. \emph{Nano Letters} \textbf{2009}, \emph{9}, 1433--1441,
  \_eprint: https://doi.org/10.1021/nl8032697\relax
\mciteBstWouldAddEndPuncttrue
\mciteSetBstMidEndSepPunct{\mcitedefaultmidpunct}
{\mcitedefaultendpunct}{\mcitedefaultseppunct}\relax
\EndOfBibitem
\bibitem[Rivera \latin{et~al.}(2015)Rivera, Schaibley, Jones, Ross, Wu,
  Aivazian, Klement, Seyler, Clark, Ghimire, Yan, Mandrus, Yao, and
  Xu]{rivera_observation_2015}
Rivera,~P.; Schaibley,~J.~R.; Jones,~A.~M.; Ross,~J.~S.; Wu,~S.; Aivazian,~G.;
  Klement,~P.; Seyler,~K.; Clark,~G.; Ghimire,~N.~J.; Yan,~J.; Mandrus,~D.~G.;
  Yao,~W.; Xu,~X. Observation of long-lived interlayer excitons in monolayer
  {MoSe2}–{WSe2} heterostructures. \emph{Nature Communications}
  \textbf{2015}, \emph{6}, 6242, Publisher: Nature Publishing Group\relax
\mciteBstWouldAddEndPuncttrue
\mciteSetBstMidEndSepPunct{\mcitedefaultmidpunct}
{\mcitedefaultendpunct}{\mcitedefaultseppunct}\relax
\EndOfBibitem
\bibitem[Rivera \latin{et~al.}(2016)Rivera, Seyler, Yu, Schaibley, Yan,
  Mandrus, Yao, and Xu]{rivera_valley-polarized_2016}
Rivera,~P.; Seyler,~K.~L.; Yu,~H.; Schaibley,~J.~R.; Yan,~J.; Mandrus,~D.~G.;
  Yao,~W.; Xu,~X. Valley-polarized exciton dynamics in a {2D} semiconductor
  heterostructure. \emph{Science} \textbf{2016}, \emph{351}, 688--691,
  Publisher: American Association for the Advancement of Science\relax
\mciteBstWouldAddEndPuncttrue
\mciteSetBstMidEndSepPunct{\mcitedefaultmidpunct}
{\mcitedefaultendpunct}{\mcitedefaultseppunct}\relax
\EndOfBibitem
\bibitem[Seyler \latin{et~al.}(2019)Seyler, Rivera, Yu, Wilson, Ray, Mandrus,
  Yan, Yao, and Xu]{seyler_signatures_2019}
Seyler,~K.~L.; Rivera,~P.; Yu,~H.; Wilson,~N.~P.; Ray,~E.~L.; Mandrus,~D.~G.;
  Yan,~J.; Yao,~W.; Xu,~X. Signatures of moiré-trapped valley excitons in
  {MoSe2}/{WSe2} heterobilayers. \emph{Nature} \textbf{2019}, \emph{567},
  66--70, Publisher: Nature Publishing Group\relax
\mciteBstWouldAddEndPuncttrue
\mciteSetBstMidEndSepPunct{\mcitedefaultmidpunct}
{\mcitedefaultendpunct}{\mcitedefaultseppunct}\relax
\EndOfBibitem
\bibitem[Shabani \latin{et~al.}(2021)Shabani, Halbertal, Wu, Chen, Liu, Hone,
  Yao, Basov, Zhu, and Pasupathy]{shabani_deep_2021}
Shabani,~S.; Halbertal,~D.; Wu,~W.; Chen,~M.; Liu,~S.; Hone,~J.; Yao,~W.;
  Basov,~D.~N.; Zhu,~X.; Pasupathy,~A.~N. Deep moiré potentials in twisted
  transition metal dichalcogenide bilayers. \emph{Nature Physics}
  \textbf{2021}, \emph{17}, 720--725, Publisher: Nature Publishing Group\relax
\mciteBstWouldAddEndPuncttrue
\mciteSetBstMidEndSepPunct{\mcitedefaultmidpunct}
{\mcitedefaultendpunct}{\mcitedefaultseppunct}\relax
\EndOfBibitem
\bibitem[Pendharkar \latin{et~al.}(2024)Pendharkar, Tran, Zaborski, Finney,
  Sharpe, Kamat, Kalantre, Hocking, Bittner, Watanabe, Taniguchi, Pittenger,
  Newcomb, Kastner, Mannix, and Goldhaber-Gordon]{pendharkar_torsional_2024}
Pendharkar,~M. \latin{et~al.}  Torsional force microscopy of van der {Waals}
  moirés and atomic lattices. \emph{Proceedings of the National Academy of
  Sciences} \textbf{2024}, \emph{121}, e2314083121, Publisher: Proceedings of
  the National Academy of Sciences\relax
\mciteBstWouldAddEndPuncttrue
\mciteSetBstMidEndSepPunct{\mcitedefaultmidpunct}
{\mcitedefaultendpunct}{\mcitedefaultseppunct}\relax
\EndOfBibitem
\bibitem[Li \latin{et~al.}(2013)Li, Rao, Mak, You, Wang, Dean, and
  Heinz]{li_probing_2013}
Li,~Y.; Rao,~Y.; Mak,~K.~F.; You,~Y.; Wang,~S.; Dean,~C.~R.; Heinz,~T.~F.
  Probing symmetry properties of few-layer {MoS2} and h-{BN} by optical
  second-harmonic generation. \emph{Nano letters} \textbf{2013}, \emph{13},
  3329--3333, Publisher: ACS Publications\relax
\mciteBstWouldAddEndPuncttrue
\mciteSetBstMidEndSepPunct{\mcitedefaultmidpunct}
{\mcitedefaultendpunct}{\mcitedefaultseppunct}\relax
\EndOfBibitem
\bibitem[Kim \latin{et~al.}(2019)Kim, Fröch, Gardner, Li, Aharonovich, and
  Solntsev]{kim_second-harmonic_2019}
Kim,~S.; Fröch,~J.~E.; Gardner,~A.; Li,~C.; Aharonovich,~I.; Solntsev,~A.~S.
  Second-harmonic generation in multilayer hexagonal boron nitride flakes.
  \emph{Optics letters} \textbf{2019}, \emph{44}, 5792--5795, Publisher: Optica
  Publishing Group\relax
\mciteBstWouldAddEndPuncttrue
\mciteSetBstMidEndSepPunct{\mcitedefaultmidpunct}
{\mcitedefaultendpunct}{\mcitedefaultseppunct}\relax
\EndOfBibitem
\bibitem[Shan \latin{et~al.}(2018)Shan, Li, Huang, Tong, Yao, Liu, and
  Wu]{shan_stacking_2018}
Shan,~Y.; Li,~Y.; Huang,~D.; Tong,~Q.; Yao,~W.; Liu,~W.-T.; Wu,~S. Stacking
  symmetry governed second harmonic generation in graphene trilayers.
  \emph{Science Advances} \textbf{2018}, \emph{4}, eaat0074, \_eprint:
  https://www.science.org/doi/pdf/10.1126/sciadv.aat0074\relax
\mciteBstWouldAddEndPuncttrue
\mciteSetBstMidEndSepPunct{\mcitedefaultmidpunct}
{\mcitedefaultendpunct}{\mcitedefaultseppunct}\relax
\EndOfBibitem
\bibitem[HEINZ(1991)]{ponath_chapter_1991}
HEINZ,~T.~F. In \emph{Nonlinear {Surface} {Electromagnetic} {Phenomena}};
  PONATH,~H.-E., STEGEMAN,~G.~I., Eds.; Modern {Problems} in {Condensed}
  {Matter} {Sciences}; Elsevier, 1991; Vol.~29; pp 353--416, ISSN:
  0167-7837\relax
\mciteBstWouldAddEndPuncttrue
\mciteSetBstMidEndSepPunct{\mcitedefaultmidpunct}
{\mcitedefaultendpunct}{\mcitedefaultseppunct}\relax
\EndOfBibitem
\bibitem[Grüneis \latin{et~al.}(2003)Grüneis, Saito, Samsonidze, Kimura,
  Pimenta, Jorio, Souza~Filho, Dresselhaus, and
  Dresselhaus]{gruneis_inhomogeneous_2003}
Grüneis,~A.; Saito,~R.; Samsonidze,~G.~G.; Kimura,~T.; Pimenta,~M.; Jorio,~A.;
  Souza~Filho,~A.; Dresselhaus,~G.; Dresselhaus,~M. Inhomogeneous optical
  absorption around the {K} point in graphite and carbon nanotubes.
  \emph{Physical Review B} \textbf{2003}, \emph{67}, 165402, Publisher:
  APS\relax
\mciteBstWouldAddEndPuncttrue
\mciteSetBstMidEndSepPunct{\mcitedefaultmidpunct}
{\mcitedefaultendpunct}{\mcitedefaultseppunct}\relax
\EndOfBibitem
\bibitem[Krishna~Kumar \latin{et~al.}(2017)Krishna~Kumar, Chen, Auton,
  Mishchenko, Bandurin, Morozov, Cao, Khestanova, Ben~Shalom, Kretinin,
  Novoselov, Eaves, Grigorieva, Ponomarenko, Fal’ko, and
  Geim]{krishna_kumar_high-temperature_2017}
Krishna~Kumar,~R. \latin{et~al.}  High-temperature quantum oscillations caused
  by recurring {Bloch} states in graphene superlattices. \emph{Science}
  \textbf{2017}, \emph{357}, 181--184, Publisher: American Association for the
  Advancement of Science\relax
\mciteBstWouldAddEndPuncttrue
\mciteSetBstMidEndSepPunct{\mcitedefaultmidpunct}
{\mcitedefaultendpunct}{\mcitedefaultseppunct}\relax
\EndOfBibitem
\bibitem[Sumaiya \latin{et~al.}(2022)Sumaiya, Liu, and
  Baykara]{sumaiya_true_2022}
Sumaiya,~S.~A.; Liu,~J.; Baykara,~M.~Z. True {Atomic}-{Resolution} {Surface}
  {Imaging} and {Manipulation} under {Ambient} {Conditions} via {Conductive}
  {Atomic} {Force} {Microscopy}. \emph{ACS Nano} \textbf{2022}, \emph{16},
  20086--20093, Publisher: American Chemical Society\relax
\mciteBstWouldAddEndPuncttrue
\mciteSetBstMidEndSepPunct{\mcitedefaultmidpunct}
{\mcitedefaultendpunct}{\mcitedefaultseppunct}\relax
\EndOfBibitem
\bibitem[Hu \latin{et~al.}(2023)Hu, Tan, Al~Ezzi, Chattopadhyay, Gou, Zheng,
  Wang, Chen, Thottathil, Luo, Watanabe, Taniguchi, Wee, Adam, and
  Ariando]{hu_controlled_2023}
Hu,~J.; Tan,~J.; Al~Ezzi,~M.~M.; Chattopadhyay,~U.; Gou,~J.; Zheng,~Y.;
  Wang,~Z.; Chen,~J.; Thottathil,~R.; Luo,~J.; Watanabe,~K.; Taniguchi,~T.;
  Wee,~A. T.~S.; Adam,~S.; Ariando,~A. Controlled alignment of supermoiré
  lattice in double-aligned graphene heterostructures. \emph{Nature
  Communications} \textbf{2023}, \emph{14}, 1--8, Number: 1 Publisher: Nature
  Publishing Group\relax
\mciteBstWouldAddEndPuncttrue
\mciteSetBstMidEndSepPunct{\mcitedefaultmidpunct}
{\mcitedefaultendpunct}{\mcitedefaultseppunct}\relax
\EndOfBibitem
\bibitem[Kamat \latin{et~al.}(2024)Kamat, Sharpe, Pendharkar, Hu, Tran,
  Zaborski~Jr., Hocking, Finney, Watanabe, Taniguchi, Kastner, Mannix, , Heinz,
  and Goldhaber-Gordon]{data_repo}
Kamat,~R.~V.; Sharpe,~A.~L.; Pendharkar,~M.; Hu,~J.; Tran,~S.~J.;
  Zaborski~Jr.,~G.; Hocking,~M.; Finney,~J.; Watanabe,~K.; Taniguchi,~T.;
  Kastner,~M.~A.; Mannix,~A.~J.; ; Heinz,~T.; Goldhaber-Gordon,~D. Data for:
  Deterministic fabrication of graphene hexagonal boron nitride moire
  superlattices. 2024; \url{https://doi.org/10.25740/np508hx1441}, Stanford
  Digital Repository. https://doi.org/10.25740/np508hx1441. Deposited 24 May
  2024.\relax
\mciteBstWouldAddEndPunctfalse
\mciteSetBstMidEndSepPunct{\mcitedefaultmidpunct}
{}{\mcitedefaultseppunct}\relax
\EndOfBibitem
\bibitem[Blakslee \latin{et~al.}(2003)Blakslee, Proctor, Seldin, Spence, and
  Weng]{blakslee_elastic_2003}
Blakslee,~O.~L.; Proctor,~D.~G.; Seldin,~E.~J.; Spence,~G.~B.; Weng,~T. Elastic
  {Constants} of {Compression}‐{Annealed} {Pyrolytic} {Graphite}.
  \emph{Journal of Applied Physics} \textbf{2003}, \emph{41}, 3373--3382\relax
\mciteBstWouldAddEndPuncttrue
\mciteSetBstMidEndSepPunct{\mcitedefaultmidpunct}
{\mcitedefaultendpunct}{\mcitedefaultseppunct}\relax
\EndOfBibitem
\bibitem[Seldin and Nezbeda(2003)Seldin, and Nezbeda]{seldin_elastic_2003}
Seldin,~E.~J.; Nezbeda,~C.~W. Elastic {Constants} and {Electron}‐{Microscope}
  {Observations} of {Neutron}‐{Irradiated} {Compression}‐{Annealed}
  {Pyrolytic} and {Single}‐{Crystal} {Graphite}. \emph{Journal of Applied
  Physics} \textbf{2003}, \emph{41}, 3389--3400\relax
\mciteBstWouldAddEndPuncttrue
\mciteSetBstMidEndSepPunct{\mcitedefaultmidpunct}
{\mcitedefaultendpunct}{\mcitedefaultseppunct}\relax
\EndOfBibitem
\bibitem[Bosak \latin{et~al.}(2007)Bosak, Krisch, Mohr, Maultzsch, and
  Thomsen]{bosak_elasticity_2007}
Bosak,~A.; Krisch,~M.; Mohr,~M.; Maultzsch,~J.; Thomsen,~C. Elasticity of
  single-crystalline graphite: {Inelastic} x-ray scattering study.
  \emph{Physical Review B} \textbf{2007}, \emph{75}, 153408, Publisher:
  American Physical Society\relax
\mciteBstWouldAddEndPuncttrue
\mciteSetBstMidEndSepPunct{\mcitedefaultmidpunct}
{\mcitedefaultendpunct}{\mcitedefaultseppunct}\relax
\EndOfBibitem
\end{mcitethebibliography}

\end{document}